\documentclass[journal,twocolumn]{IEEEtran}

\usepackage[cmex10]{amsmath}
\usepackage{amssymb}
\usepackage{cite}
\usepackage{url}
\usepackage{graphicx}
\usepackage{array,color}
\usepackage{multirow}
\usepackage{amsmath}
\usepackage{stfloats}
\usepackage{graphicx}
\usepackage{subfigure}
\usepackage{tabularx}
\usepackage{epsfig,color,balance,cite}
\usepackage{setspace}
\usepackage{bm}
\usepackage{textcomp}
\usepackage{algorithmic}
\usepackage{algorithm}
\usepackage{comment}
\usepackage{caption}
\usepackage{graphicx}
\usepackage{setspace}
\usepackage{diagbox}
\usepackage{float}
\usepackage{epstopdf}
\usepackage{caption}
\usepackage{subcaption}
\usepackage{booktabs}
\usepackage{array}
\usepackage{makecell}
\newcolumntype{Y}{>{\centering\arraybackslash}X}
\UseRawInputEncoding

\usepackage[cmex10]{amsmath}
\usepackage{amssymb}
\usepackage{cite}
\usepackage{graphicx}
\usepackage{array,color}
\usepackage{multirow}
\usepackage{amsmath}
\usepackage{stfloats}
\usepackage{graphicx}
\usepackage{subfigure}
\usepackage{tabularx}
\usepackage{epsfig,epsf,color,balance,cite}
\usepackage{setspace}
\usepackage{bm}
\usepackage{textcomp}
\usepackage{algorithmic}
\usepackage{algorithm}
\usepackage{caption} 

\usepackage{comment}
\usepackage{subfigure}

\usepackage{caption}
\usepackage{graphicx}
\usepackage{setspace}
\usepackage{diagbox}

\usepackage{multirow}
\usepackage{graphicx}
\allowdisplaybreaks[4]
\makeatletter

\newcommand{\Rmnum}[1]{\expandafter\@slowromancap\romannumeral #1@}
\makeatother
\usepackage{booktabs}
\usepackage{amsmath}
\newtheorem{Theorem}{Theorem }

\begin{document}
	
	\title{A Cyclic Shift Embedded Pilot based Channel Estimation for Multi-User MIMO-OTFS systems with fractional delay and Doppler}
	
	\author{Ruizhe Wang, Hong Ren, Cunhua Pan, Ruisong Weng, and Jiangzhou Wang, \IEEEmembership{Fellow, IEEE}
		\thanks{R. Wang, H. Ren, C. Pan, R. Weng and J. Wang are with National Mobile Communications Research Laboratory, Southeast University, Nanjing, China. (e-mail: {rzw, hren, cpan, ruisong\_weng, j.z.wang}@seu.edu.cn).

		\emph{Corresponding author: Hong Ren}}
	}
	\maketitle


\begin{abstract}
Orthogonal time frequency space (OTFS) modulation has been proposed to meet the demand for reliable communication in high-mobility scenarios for future wireless networks. However, in multi-user OTFS systems, conventional embedded pilot schemes require independent pilot allocation for each user, leading to linearly increasing pilot overhead.
To address these issues, in this paper, we investigate the uplink channel estimation and pilot design for multi-user multiple-input multiple-output (MIMO)-OTFS  systems. 
We propose a multi-dimensional decomposition-based channel estimation algorithm. 
Specifically, the proposed algorithm first estimates the angles of arrivals (AoAs) via subspace decomposition-based method. A spatial projection matrix, constructed from the estimated AOAs, decouples the received signal by propagation path subspace, effectively mitigating inter-path interference.
The remaining fractional delay and Doppler can be obtained by a compressed sensing (CS)-based off-grid channel estimation method.
Furthermore, to reduce the pilot overhead in multi-user OTFS systems, this paper proposes a novel cyclic shift embedded pilot (CSEP) structure, which can reuse users through cyclic shift-orthogonality of Zadoff-Chu (ZC) sequences. 
Compared with conventional embedded pilot structures, the CSEP structure can save over 30\% of pilot overhead. 
Finally, an imporved channel estimation method based on the CSEP structure is proposed.
Simulation results demonstrate that it achieves superior performance in channel estimation. 
Moreover, the proposed CSEP structure and channel estimation algorithm achieve a favorable balance between computational complexity, estimation accuracy, and bit error rate (BER) performance.
\end{abstract}


\IEEEpeerreviewmaketitle
%
\section{Introduction}
Future sixth-generation (6G) wireless networks are envisioned to provide global ubiquitous coverage for various mobile terminals, including high speed vehicles, unmanned aerial vehicles (UAVs), high speed trains and low-earth-orbit (LEO) satellites, etc \cite{6g}. 
However, the Doppler spread induced by high-speed mobility disrupts the orthogonality among subcarriers in orthogonal frequency division multiplexing (OFDM) systems.
This causes spectral overlap of adjacent subcarriers, leading to severe inter-carrier interference (ICI), thereby degrading communication reliability \cite{otfstutorial}. 
Consequently, ensuring reliable communications in high-mobility scenarios has become one of the critical challenges for future 6G wireless networks.

Fortunately, orthogonal time-frequency space (OTFS) modulation has emerged as a promising solution to this challenge \cite{otfs,otfs2}. OTFS is a two-dimensional modulation technique that exploits the sparse representation of dominant channel scattering paths in the delay-Doppler (DD) domain. Compared with conventional frequency-domain channel, the DD domain channel exhibits greater stability in high-mobility scenarios. Moreover, the implementation of OTFS can be made compatible with the existing OFDM systems by integrating inverse symplectic Fourier transform (ISFFT) preprocessing  and symplectic Fourier transform (SFFT) postprocessing modules, thereby reducing deployment costs while retaining the advantages of OFDM \cite{ofdmotfs}.

Accurate channel estimation is the key to achieving reliable communication promised by OTFS systems. One of the key challenges in OTFS channel estimation is the estimation of fractional delay and Doppler. The channel sparsity of the DD channel is degraded with fractional delay and Doppler, which poses significant challenges for channel reconstruction. 
In \cite{embeddedpilot}, an embedded pilot structure was proposed, which employs a central impulse pilot surrounded by guard symbols to mitigate interference. The channel estimation is converted to a simple impulse detection in the DD domain. However, this approach merely captures a sampled representation of center DD domain channel rather than the physical channel. 
The work in \cite{sblotfs} addressed the channel estimation problem via the sparse Bayesian learning (SBL) framework. 
A grid-evolution approach was applied, which linearly approximates the fractional Doppler via a first-order Taylor expansion, thereby enabling iterative recovery of accurate channel. To reduce the computational complexity, \cite{sblotfs3} introduced a fast Bayesian compressive sensing (FBCS) algorithm utilizing a Laplacian Scale Mixture (LSM) prior. The work in \cite{shan} further refined the grid-evolution method by adopting a Student's t-distribution as the prior to reduce the computational complexity. Nevertheless, the computational complexity of the SBL-based method still increases considerably with the size of the channel, especially in MIMO-OTFS systems. 

As a lower-complexity alternative compressed sensing (CS) algorithm, orthogonal matching pursuit (OMP) has been also applied widely in the OTFS channel estimation. \cite{shen} proposed a 3D structured OMP algorithm for MIMO-OTFS channel estimation. However, this method relies on the assumption that the delays and Dopplers are on-grid. In practical scenarios, this assumption causes a significant model mismatch, leading to severe degradation in estimation accuracy. To address the off-grid issue, \cite{dingshi} proposed an SOMP-based algorithm that estimates the fractional Doppler by leveraging the amplitude profile of the DD domain channel. A notable drawback of this approach is its reliance on redundant pilots, which results in degraded spectral efficiency (SE).  
By exploiting the sparsity in the four-dimensional delay-Doppler-angle domain, \cite{multidim} proposed a 4D-OMP-based algorithm. 
However, multi-dimensional compressed sensing involves high-dimensional measurement matrices, leading to high computational complexity.

Achieving accurate channel estimation with low pilot overhead in multi-user OTFS systems also remains a challenging problem.
In \cite{tfmultiaccess}, the orthogonal resources are allocated to the users in the time-frequency (TF) domain.
\cite{multiaccess,multiaccess5} adopted the method of allocating independent pilot regions in the Doppler domain for different users. 
The results in \cite{multiaccess7} revealed that the DD-domain multiple-access (MA) scheme employing rectangular pulses achieves superior system and spectral efficiency compared to TF-domain MA and guard-band-based MA schemes.​
A synchronization technique for uplink multi-user OTFS systems was presented in \cite{multiaccess2}, in which multi-user pilots were designed as cyclic-prefixed sequences arranged along the delay or Doppler domain. Time and carrier frequency offsets are estimated using a maximum likelihood search. ​A key limitation of this scheme, however, is its inability to handle fractional delay and Doppler. 
In \cite{multiaccess6}, the transmitted signal for each user is confined to a sub-region of the time-frequency grid by assigning uniformly spaced DD resource blocks to each user.
The work \cite{multiaccess3} proposed a scheme to separate multiple users by inserting guard intervals in the DD domain and employed a 3D Newtonized OMP (NOMP) algorithm for channel estimation. However,  the required guard interval length increases with the number of users, which still leads to a significant pilot overhead. 
To further improve the utilization of the delay-Doppler resource grid, the work \cite{multiaccess4} employed the superimposed pilot schemes and proposed a message passing (MP)-based channel estimation algorithm to separate pilot signals from data signals. However, this scheme suffers from a significant degradation in both channel estimation accuracy and bit error rate (BER) performance.
Therefore, it is imperative to develop a pilot scheme for multi-user OTFS systems that can achieve a favorable trade-off between pilot overhead and system performance.

​Motivated by the challenges discussed above, in this paper, we investigate the pilot design and channel estimation for multi-user MIMO-OTFS  systems with the consideration of fractional delay and Doppler. The main contributions of this paper are summarized as follows:

\begin{itemize}
	\item We analyze the input-output relationship for multi-user MIMO-OTFS systems with the consideration of the practical fractional delay and Doppler. A channel estimation algorithm based on the CS algorithm and subspace decomposition is developed. Unlike general 3D CS methods, the proposed algorithm employs a multi-dimensional decomposition approach. By projecting signals using the estimated angle information, the interference among multiple paths in the DD domain is mitigated, thereby effectively enhancing the estimation accuracy. Compared to the SBL-based method, the proposed algorithm has lower computational complexity.
	\item To reduce the pilot overhead in multi-user scenarios, we first propose a cyclic shift embedded pilot (CSEP)  structure based on Zadoff-Chu (ZC) sequences. This pilot structure reduces the delay-domain pilot overhead by $KM_g$ when serving $K$ users, where $M_g$ denotes the guard interval along the delay domain. The proposed CSEP structure can achieve over 30\% overhead savings than conventional pilot structures when $K$ is large.
	An improved channel estimation algorithm is then developed for the proposed CSEP structure.
	\item Although the CSEP pilot structure introduces a certain level of inter-user interference, it provides each user with a larger number of usable pilot symbols compared to the conventional pilot structure, which increase the noise-resistance. Simulation results demonstrate that the proposed CSEP structure and channel estimation algorithm outperform the benchmark schemes. Furthermore, comparisons of the BER and SE performance between OFDM and OTFS are included, which substantiate the necessity of employing OTFS in high-mobility environments.
\end{itemize}

\section{MIMO-OTFS Multiuser System Model}\label{systemmodel}
\subsection{MIMO-OTFS Signal Model}
\begin{figure*}[ht]
	\begin{minipage}[ht]{1\linewidth}
		\centering
		\includegraphics[width=1.0\linewidth]{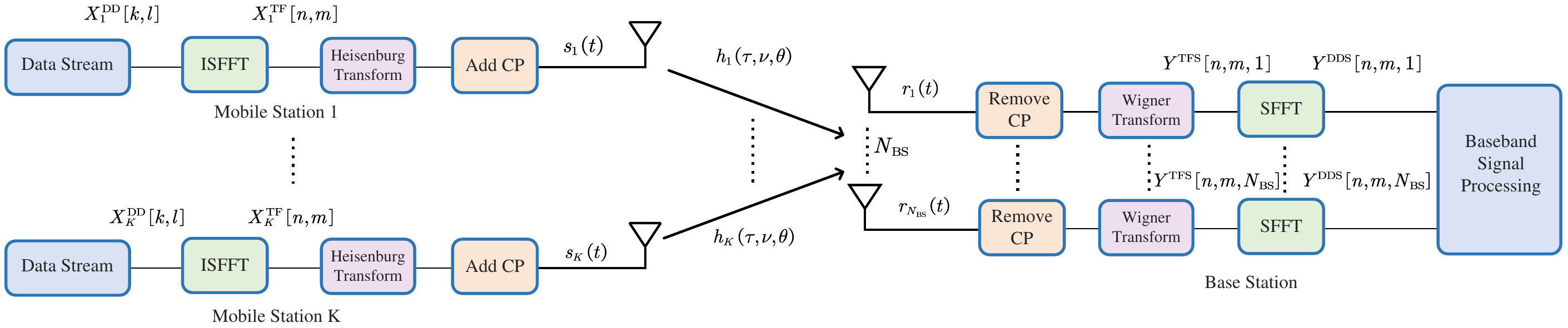}
		\caption{An example of MIMO-OTFS multiuser systems.}
		\label{systemmodel}
	\end{minipage}%
	\hfill
\end{figure*}
Consider an MIMO-OTFS multiuser system shown in Fig.~\ref{systemmodel}. The $K$ mobile stations (MSs) are equipped with single antenna, and the base station (BS) is equipped with  a uniform
linear array (ULA) of $N_{\mathrm{BS}}$ antennas. The cyclic-prefix (CP)-OTFS is considered. The number of intervals per OTFS frame and the number of subcarriers are $N$ and $M$, respectively, which corresponds to a discrete time-frequency (TF) grid $\Lambda =\left\{ \left( nT_{\mathrm{sym}},m\Delta f \right)\right\} $, where $n=0,1,\cdots ,N-1$, and $m=0,1,\cdots ,M-1 $, respectively, $T_{\mathrm{sym}}$ and $\Delta f$ are the duration per symbol with CP and the subcarrier spacing, respectively. 
The corresponding discrete delay-Doppler (DD) grid is defined as $\varGamma =\left\{ \left( \frac{k}{NT_{\mathrm{sym}}},\frac{l}{M\Delta f} \right)\right\}$, where $k=-\lceil \frac{N}{2} \rceil ,-\lceil \frac{N}{2} \rceil +1,..., \lceil \frac{N}{2} \rceil$, and $l=0,1,...,M-1  $, respectively. Denote the delay-Doppler symbol of the $u$-th mobile station as $X_{u}^{\mathrm{DD}}\left[ k,l \right] $. At the $u$-th mobile station, the delay-Doppler domain symbols $X_{u}^{\mathrm{DD}}\left[ k,l \right] $ are firstly mapped into the time-frequency domain symbols $X_{u}^{\mathrm{TF}}\left[ n,m \right] $ via inverse Symplectic Finite Fourier Transform (ISFFT):
\begin{equation}\label{XTFu}
	X_{u}^{\mathrm{TF}}\left[ n,m \right] =\frac{1}{\sqrt{NM}}\sum_{k=-\lceil N/2 \rceil}^{\lceil N/2 \rceil}{\sum_{l=0}^{M-1}{X_{u}^{\mathrm{DD}}\left[ k,l \right] e^{-j2\pi \left( \frac{ml}{M}-\frac{nk}{N} \right)}}},
\end{equation}
where $T$ denotes the symbol duration without CP. We have $T=\frac{1}{\Delta f}$ and $T=\frac{M}{M+M_{\mathrm{CP}}}T_{\mathrm{sym}}$, where $M_{\mathrm{CP}}$ denotes the length of CP.
Then, the TF domain symbols are transformed into transmitted signal with rectangular waveform \cite{dingshi} as
\begin{equation}\label{sut}
	\begin{aligned}
		s_{u}\left( t \right) &=\sum_{n=0}^{N-1}{\sum_{m=0}^{M-1}{X_{u}^{\mathrm{TF}}\left[ n,m \right] e^{j2\pi m\Delta f\left( t-\frac{M_{\mathrm{CP}}T}{M}-nT_{\mathrm{sym}} \right)}}}\\
		&\qquad\qquad\qquad\qquad\qquad\qquad\qquad \times  g_{\mathrm{tx}}\left( t-nT_{\mathrm{sym}} \right),
	\end{aligned}
\end{equation}
where the rectangular function $ g_{\mathrm{tx}}\left( t \right)$ is defined as 
\begin{equation}
	g_{\mathrm{tx}}\left( t \right) =\begin{cases}
		\frac{1}{\sqrt{T}},\quad 0\le t\le T_{\mathrm{sym}}\\
		0,\qquad \mathrm{otherwise}.\\
	\end{cases}
\end{equation}
After multicarrier modulation, $K$ MSs transmit their signals through the multipath time-variant channel, and the received signal $	\mathbf{r}\left( t \right)$ at the BS is given by
\begin{equation}\label{rt}
	\begin{aligned}
	\mathbf{r}\left( t \right) &=\sum_u{}\iiint{}\mathbf{a}_{N_{\mathrm{BS}}}\left( \theta \right) h_u\left( \tau ,\nu ,\theta \right) \\
	&\qquad\times e^{j2\pi \nu \left( t-\tau \right)}s_u\left( t-\tau \right) \mathrm{d}\tau \mathrm{d}\nu \mathrm{d}\theta +\mathbf{n}\left( t \right)  ,
	\end{aligned}
\end{equation}
where $h_u\left( \tau ,\nu ,\theta \right) $ denotes the complex channel impulse response in the delay-Doppler-space domain for the $u$-th MS \cite{jakes1994microwave},  with
$\tau $, $\nu$ and $\theta $ representing the delay, the Doppler and the arrival of angle (AOA), respectively. 
For a simple ULA, the array steering vector $\mathbf{a}_X\left( \theta \right) $ is defined as $\mathbf{a}_X\left( \theta \right) =\left[ 1, e^{j2\pi \frac{d}{\lambda}\cos \theta},..., e^{j2\pi \frac{d}{\lambda}\left( X-1 \right) \cos \theta} \right] ^{\mathrm{T}}$. 
$\mathbf{n}\left( t \right) $ denotes the additive white Gaussian noise. 
The delay-Doppler-space domain channel impulse response for the $u$-th MS $h_u\left( \tau ,\nu ,\theta \right) $ is given by 
\begin{equation}\label{hDDu}
	h_u\left( \tau ,\nu ,\theta \right) =\sum_i{\alpha _{u,i}}\delta \left( \tau -\tau _{u,i} \right) \delta \left( \nu -\nu _{u,i} \right) \delta \left( \theta -\theta _{u,i} \right)  ,
\end{equation}
where ${\alpha}_{u,i}$, ${\tau}_{u,i}$, $\nu _{u,i}$ and $\theta _{u,i}$ denote the complex channel gain, the delay, the Doppler and the AOA for the $u$-th MS of the $i$-th propagation path, respectively. The delay and Doppler taps for the $u$-th MS of the $i$-th path can be denoted as 
\begin{equation}\label{frackili}
	\tau _{u,i}=\frac{l_{u,i}+\iota _{u,i}}{M\Delta f}, \nu _{u,i}=\frac{k_{u,i}+\kappa _{u,i}}{NT_{\mathrm{sym}}},
\end{equation}
where $l_{u,i}$ and $k_{u,i}$ are the integer index of delay tap and Doppler tap for the $u$-th MS of the $i$-th propagation path, and $\iota _{u,i}\in \left( -\frac{1}{2},\frac{1}{2} \right]$ and $\kappa _{u,i}\in \left( -\frac{1}{2},\frac{1}{2} \right]$ are the fractional delay and Doppler for the $u$-th MS of the $i$-th propagation path, respectively. 
At the BS, the received signal at the $n_{\mathrm{BS}}$-th antenna is processed through a match filter as
\begin{equation}\label{Atf}
	\begin{aligned}
		Y\left( t,f,n_{\mathrm{BS}} \right) &=A_{g_{\mathrm{rx}},r}\left( t,f,n_{\mathrm{BS}} \right) 
		\\
		&\triangleq \int{e^{-j2\pi f\left( t^{\prime}-t-T_{\mathrm{CP}} \right)}g_{\mathrm{rx}}^{\ast}\left( t^{\prime}-t \right) r_{n_{\mathrm{BS}}}\left( t^{\prime} \right) \mathrm{d}t^{\prime}},
	\end{aligned}
\end{equation}
where $g_{\mathrm{rx}}\left( t \right) $ is defined as 
\begin{equation}
	g_{\mathrm{rx}}\left( t \right) =\begin{cases}
		\frac{1}{\sqrt{T}},\quad\frac{M_{\mathrm{CP}}T}{M}\le t\le T_{\mathrm{sym}}\\
		0,\qquad\mathrm{otherwise}.\\
	\end{cases}
\end{equation}
Then, the received symbols in the time-frequency-space (TFS) domain can be extracted by sampling  $Y\left( t,f,n_{\mathrm{BS}} \right)$ as 
\begin{equation}\label{YTFSsample}
	Y^{\mathrm{TFS}}\left[ n,m,n_{\mathrm{BS}} \right] =Y\left( nT_{\mathrm{sym}},m\Delta f,n_{\mathrm{BS}} \right) .
\end{equation}
Next, the sympletic finite Fourier transform (SFFT) maps the TFS domain symbols $Y^{\mathrm{TFS}}\left[ n,m,n_{\mathrm{BS}} \right] $ to the DD domain symbols $Y^{\mathrm{DDS}}\left[ k,l,n_{\mathrm{BS}} \right] $ as
\begin{equation}\label{YDDS}
	\begin{aligned}
		&Y^{\mathrm{DDS}}\left[ k,l,n_{\mathrm{BS}} \right]\\ &\qquad=\frac{1}{\sqrt{NM}}\sum_{n=0}^{N-1}{\sum_{m=0}^{M-1}{Y^{\mathrm{TFS}}\left[ n,m,n_{\mathrm{BS}} \right] e^{-j2\pi \left( \frac{nk}{N}-\frac{ml}{M} \right)}}}.
	\end{aligned}
\end{equation}
The input-output relationship in the delay-Doppler-spatial (DDS) domain can be characterized by following theorem.
\begin{Theorem}\label{theorem1}
	The input-output relationship of the MIMO-OTFS multiuser system can be expressed as
	\begin{equation}\label{YDDderivation}
			\begin{aligned}
				&Y_{}^{\mathrm{DDS}}\left[ k,l,n_{\mathrm{BS}} \right]\\
				&\quad =\sum_{u=1}^K{}\sum_{i=0}^{P-1}{}\sum_{k^{\prime}=-\lceil N/2 \rceil}^{\lceil N/2 \rceil}{\sum_{l^{\prime}=0}^{M-1}{}\Phi _{u,i}\left( l \right) h_{u,i}\left[ k^{\prime},l^{\prime},n_{\mathrm{BS}} \right]}\\
				&\qquad \times X_{u}^{\mathrm{DD}}\left[ \left[ k-k^{\prime} \right] _N,\left[ l-l^{\prime} \right] _M \right] +Z^{\mathrm{DDS}}\left[ k,l,n_{\mathrm{BS}} \right] ,\\
			\end{aligned}
	\end{equation}
	where $Z^{\mathrm{DDS}}\left[ k,l,n_{\mathrm{BS}} \right]$ is the additive noise and 
	\begin{equation}\label{hphiXi}
		\begin{aligned}
			h_{u,i}\left[ k^{\prime},l^{\prime},n_{\mathrm{BS}} \right] &=\alpha _{u,i}\left[ \mathbf{a}_{N_{\mathrm{BS}}}\left( \theta _{u,i} \right) \right] _{n_{\mathrm{BS}}}\Xi _N\left( k^\prime -k_{u,i}-\kappa _{u,i} \right) 
			\\
			&\quad\times\Xi _{M}\left( -\left(l^\prime -l_{u,i}-\iota _{u,i} \right)\right) 
			\\
			\Phi _{u,i}\left( l \right) &=e^{j2\pi \frac{\left( k_{u,i}+\kappa _{u,i} \right) \left( M_{\mathrm{CP}}+l-l_{u,i}-\iota _{u,i} \right)}{N\left( M+M_{\mathrm{CP}} \right)}}
			\\
			\Xi _X\left( x \right) &=\frac{e^{-j2\pi x}-1}{ X e^{-j\frac{2\pi}{N}x}- X },
		\end{aligned}
	\end{equation}
	where $\left[ \mathbf{a}_{N_{\mathrm{BS}}}\left( \theta _{u,i} \right) \right] _{n_{\mathrm{BS}}}$ denotes the ${n_{\mathrm{BS}}}$-th element of $\mathbf{a}_{N_{\mathrm{BS}}}\left( \theta _{u,i} \right) $.
\end{Theorem}

\textit{Proof:} The proof is provided in Appendix \ref{proofoftheorem1}. \hfill $\blacksquare$

As shown in (\ref{YDDS}), the DDS-domain signal associated with the $n_{\mathrm{BS}}$-th antenna of the BS is given by a superposition over all users and propagation paths, where for each, the 2-D convolution of the DDS domain channel $h_{u,i}\left[ k,l,n_{\mathrm{BS}} \right]$ with the respective DD domain symbols $X_{u}^{\mathrm{DD}}\left[ k,l \right] $ is then multiplied by the corresponding phase factor $\Phi _{u,i}$. 
The magnitude of the function $\Xi_X(x)$ in (\ref{hphiXi}) decreases as the Doppler domain index $k$  moves away from $k_{u,i} + \kappa_{u,\mathrm{i}}$, or as the delay domain index $l$ moves away from $l_{u,i} + \iota_{u,\mathrm{i}}$, and, as $\Xi_X(x)$ is approximately zero for $x$ far from zero, the support of $h_{u,i}$ can be treated as finite \cite{raviteja}. It is worth noting that when the resolution of the delay-Doppler plane is infinite, i.e., when $N\rightarrow \infty $ and $M\rightarrow \infty $, the fractional delays and Doppler in (\ref{frackili}) become zero, while (\ref{YDDderivation}) can be rewritten as 
\begin{equation}\label{ongridydd}
	\begin{aligned}
		&Y_{}^{\mathrm{DDS}}\left[ k,l,n_{\mathrm{BS}} \right]\\
		&\quad =\sum_{u=1}^K{}\sum_{i=0}^{P-1}{}\sum_{k^{\prime}=-\lceil N/2 \rceil}^{\lceil N/2 \rceil}{\sum_{l^{\prime}=0}^{M-1}{}e^{j2\pi \frac{k^{\prime}\left( M_{\mathrm{CP}}+l-l^{\prime} \right)}{N\left( M+M_{\mathrm{CP}} \right)}}h_{u,i}\left[ k^{\prime},l^{\prime},n_{\mathrm{BS}} \right]}\\
		&\qquad \times X_{u}^{\mathrm{DD}}\left[ \left[ k-k^{\prime} \right] _N,\left[ l-l^{\prime} \right] _M \right] +Z^{\mathrm{DDS}}\left[ k,l,n_{\mathrm{BS}} \right] .\\
	\end{aligned}
\end{equation}
In contrast to (\ref{YDDderivation}), the phase term  in $e^{j2\pi \frac{k^{\prime}\left( M_{\mathrm{CP}}+l-l^{\prime} \right)}{N\left( M+M_{\mathrm{CP}} \right)}}$ is no longer dependent on the unknown delay and Doppler to be estimated. Instead, it is solely a function of the grid indices $k^{\prime}$ and $l^{\prime}$.

\section{Conventional Pilot Structure and Multi-Dimension Decomposition Channel Estimation}\label{channelestimation}
In this section, we firstly introduce the conventional pilot structure and formulate the channel estimation problem. Then, we propose a multi-dimension decomposition channel estimation method for conventional pilot structure. 

\begin{figure}[t]
	\begin{minipage}[t]{1\linewidth}
		\centering
		\includegraphics[width=1.0\linewidth]{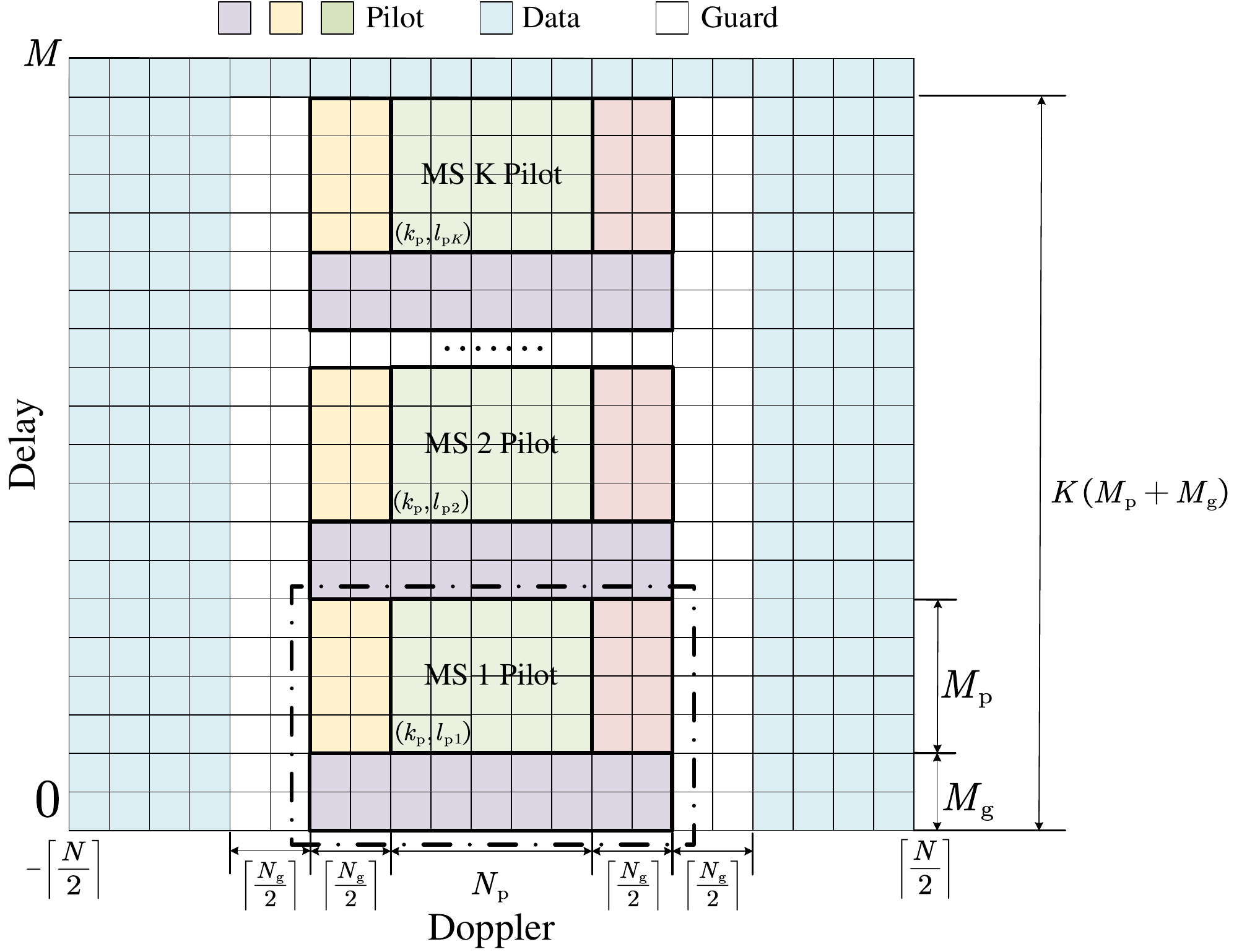}
		\caption{Conventional pilot structure for MU-OTFS systems.}
		\label{individualpilotfig}
	\end{minipage}%
	\hfill
\end{figure}

\subsection{Channel Estimation Model and Problem Formulation for Conventional Pilot Structure}\label{CEmodelandProbForm4IPS}
In this paper, the embedded pilot scheme is considered \cite{embeddedpilot}.
Fig.~\ref{individualpilotfig} shows an example of conventional pilot structure for multiuser (MU)-OTFS systems.
Denote the maximum delay and the maximum Doppler of each MS as $\tau _{\max}$ and $\nu _{\max}$, respectively.
The guard intervals (the white squares in Fig.~\ref{individualpilotfig}) are employed to eliminate the interference between the pilot symbols and the data symbols, where the guard intervals in the delay domain satisfies $M_{\mathrm{g}}> \tau _{\max}M\Delta f$ and that in Doppler domain satisfies $\lceil \frac{N_{\mathrm{g}}}{2} \rceil >\nu _{\max}NT_{\mathrm{sym}}$, respectively. 
The ranges of indices $k $ and $l $ of the delay-Doppler channel $h_{u,i}\left[ k,l \right] , \forall u,i$ are given by $\left[ -\lceil \frac{N_{\mathrm{g}}}{2} \rceil ,\lceil \frac{N_{\mathrm{g}}}{2} \rceil \right] $ and $\left[ 0,M_{\mathrm{g}} \right] $, respectively. 
As shown in Fig. 2, the dashed-dotted box outlines the complete pilot segment for MS 1. Inside this box, $\left( k_p,l_{p1} \right) $ denotes the starting position of pilot for MS 1 in the delay-Doppler plane. The green area represents the user's pilot signal, which has a length of $M_p$ in the delay domain and $N_p$ in the Doppler domain ($M_p\ge M_g$ and $N_p\ge N_g$). The yellow and purple areas are formed by repeating the last $\lceil \frac{N_{\mathrm{g}}}{2} \rceil $ columns and the first $\lceil \frac{N_{\mathrm{g}}}{2} \rceil $ columns of the green pilot array, respectively \footnote{As demonstrated in [1], if the pilot region contains only the green pilot symbols, the 2-D convolution property between the symbols and the DD-domain channel in OTFS modulation will result in the presence of zero elements in the columns of the matrix $\mathbf{X}_{u}^{\mathrm{DD}}$ in (\ref{aproYDDSnBS}).} \cite{dingshi}. Furthermore, the purple section is generated by repeating the last $M_g$ columns of the combined pilot array comprising the yellow, green, and purple areas. For the channel estimation of the $u$-th MS, the BS utilizes the DDS signal region $Y^{\mathrm{DDS}}\left[ k,l,n_{\mathrm{BS}} \right] , k=k_p, k_p+1,...,k_p+N_p-1, l=l_{pu}, l_{pu}+1,..., l_{pu}+M_p-1, \forall n_{\mathrm{BS}}$. According to the input-output relationship (\ref{YDDS}), by stacking all symbols $Y_{}^{\mathrm{DDS}}\left[ k,l,n_{\mathrm{BS}} \right]$ at the $n_{\mathrm{BS}}$-th antenna with indices $k=k_p,k_p+1,...,k_p+N_p-1, l=l_{pu},l_{pu}+1,...,l_{pu}+M_p-1$, we have
\begin{equation}\label{aproYDDSnBS}
	\mathbf{y}_{u}^{\mathrm{DD}, n_{\mathrm{BS}}}\simeq \sum_{i=0}^{P-1}{}\left( {\mathbf{X}}_{u}^{\mathrm{DD}}\odot {\mathbf{\Phi}}_{u,i} \right) \left[ \mathbf{a}_{N_{\mathrm{BS}}}\left( \theta _{u,i} \right) \right] _{n_{\mathrm{BS}}}{\mathbf{h}}_{u,i}^{\mathrm{DD}},
\end{equation}
where $\mathbf{y}_{u}^{\mathrm{DD}, n_{\mathrm{BS}}}\in \mathbb{C} ^{M_pN_p\times 1}$ with element $Y^{\mathrm{DDS}}\left[ k,l,n_{\mathrm{BS}} \right] $ of index $\left( l-l_{pu} \right) N_p+\left( k+k_p \right) +1$,
$\tilde{\mathbf{X}}_{u}^{\mathrm{DD}}\in \mathbb{C} ^{M_pN_p\times M_gN_g}$ with element $X^{\mathrm{DD}}_u\left[ \left[ k-k^{\prime} \right] _N,\left[ l-l^{\prime} \right] _M \right] $ of index $\left( \left( l-l_{pu} \right) N_p+\left( k-k_p \right) +1,\left( l^{\prime}-1 \right) N_g+\left( k^{\prime}+\lceil \frac{N_g}{2} \rceil \right) +1 \right) $, 
and ${\mathbf{\Phi}}_{u,i}$ is the phase compensation matrix for the $i$-th path of the $u$-th MS, whose elements are defined as 
\begin{equation}\label{PHIu}
	\begin{aligned}
		{\mathbf{\Phi}}_{u,i}\left( p,q \right) =\begin{cases}
			\Phi _{u,i}\left( l \right) , q\in {\mathrm{supp}\left( \mathbf{h}_{u,i} \right)} \,\, \,\,\\
			1,\qquad \ \                      \mathrm{otherwise},\\
		\end{cases}
	\end{aligned}
\end{equation}
where the indices are defined as $p=\left( l-l_{pu} \right) N_p+\left( k-k_p \right) +1$ and $q=l^{\prime}N_g+\left( k^{\prime}+\lceil \frac{N_g}{2} \rceil \right) +1$, with $k^{\prime}=-\lceil \frac{N_g}{2} \rceil ,-\lceil \frac{N_g}{2} \rceil +1,...,\lceil \frac{N_g}{2} \rceil $ and $l^{\prime}=0,1,...,M_g$, respectively. The equivalent DD domain channel ${\mathbf{h}}_{u}^{\mathrm{DD}}$ is defined as\footnote{In this work, we assume that for any two distinct paths $i$ and $j$, their delay and Doppler values are not identical simultaneously. Consequently, we have $\mathrm{supp}\left( \mathbf{h}_{u,i} \right) \cap \mathrm{supp}\left( \mathbf{h}_{u,j} \right) =\oslash $ for $i\neq j$.}
\begin{equation}
	\tilde{\mathbf{h}}_{u}^{\mathrm{DD}}\left( q \right) =\begin{cases}
		\mathbf{h}_{u,i}^{\mathrm{DD}}\left( q \right) ,q\in \bigcup\nolimits_i^{}{\mathrm{supp}\left( \mathbf{h}_{u,i} \right)}\\
		0,\qquad \quad \mathrm{otherwise}.\\
	\end{cases}
\end{equation}
By stacking all the signals in each antenna of the $u$-th MS, we have 
\begin{equation}\label{YDDSnBS}
	\begin{aligned}
\mathbf{Y}_{u}^{\mathrm{DDS}}&=\left[ \mathbf{y}_{u}^{\mathrm{DD},1},\mathbf{y}_{u}^{\mathrm{DD},2},...,\mathbf{y}_{u}^{\mathrm{DD},N_{\mathrm{BS}}} \right] ^{\mathrm{T}},
\\
&=\sum_{i=0}^{P-1}{}\mathbf{a}_{N_{\mathrm{BS}}}\left( \theta _{u,i} \right) \left( \mathbf{h}_{u,i}^{\mathrm{DD}} \right) ^{\mathrm{T}}\left( \mathbf{X}_{u}^{\mathrm{DD}}\odot \mathbf{\Phi }_{u,i} \right) ^{\mathrm{T}}
\\
&=\mathbf{H}_{u}^{\mathrm{DDS}}\left( \mathbf{X}_{u}^{\mathrm{DD}}\odot \mathbf{\Phi }_{u} \right) ^{\mathrm{T}},
	\end{aligned}
\end{equation}
where 
\begin{align}\label{H^DDS_u}
	&\mathbf{H}_{u}^{\mathrm{DDS}}\left( n_{\mathrm{BS}},q \right)\nonumber
	\\
	&\quad =\begin{cases}
		\left[ \mathbf{a}_{N_{\mathrm{BS}}}\left( \theta _{u,i} \right) \right] _{n_{\mathrm{BS}}}\mathbf{h}_{u,i}^{\mathrm{DD}}\left( q \right) ,&q\in \bigcup\nolimits_i^{}{\mathrm{supp}\left( \mathbf{h}_{u,i} \right)}\\
		0,&\mathrm{otherwise}.\\
	\end{cases}
	\\
	 &\mathbf{\Phi }_u\left( p,q \right)=\begin{cases}
		\Phi _{u,i}\left( l \right) ,&q\in \bigcup\nolimits_i^{}{}\mathrm{supp}\left( \mathbf{h}_{u,i} \right)\\
		1, &    \mathrm{otherwise},\label{PHIu}\\
	\end{cases}
\end{align}
As shown in (\ref{YDDSnBS}), the column space of  $\mathbf{Y}_{u}^{\mathrm{DDS}}$ is spanned by the set of array steering vectors $\left\{ \mathbf{a}_{N_{\mathrm{BS}}}\left( \theta _{u,i} \right) \right\} _{i=0}^{P-1}$, and the row space of matrix $\mathbf{Y}_{u}^{\mathrm{DDS}}$, on the other hand, is spanned by the set of vectors $\left\{ \left( {\mathbf{X}}_{u}^{\mathrm{DD}}\odot {\mathbf{\Phi}}_{u,i} \right) {\mathbf{h}}_{u,i}^{\mathrm{DD}} \right\} _{i=0}^{P-1}$. Therefore, the channel estimation consists of two parts: first, estimate the AoA of each path from the column space of $\mathbf{Y}_{u}^{\mathrm{DDS}}$, and second, recover the DD domain channels $\left\{ {\mathbf{h}}_{u,i}^{\mathrm{DD}} \right\} _{i=0}^{P-1}$ and estimate the Delay and Doppler of each path.

\subsection{Channel Estimation For  Conventional Pilot Structure}\label{CEforIndividual}
In this subsection, we illustrate the proposed channel estimation method for conventional pilot structure. It can be observed from (\ref{YDDSnBS}) that AoA estimation and delay and Doppler estimation are two decoupled problems. Specifically, AoA estimation is a classical harmonic retrieval and frequency estimation problem, while delay-Doppler channel estimation constitutes a CS problem whose measurement matrix ${\mathbf{\Phi}}_{u,i}$ is a dynamic grid parameterized by the delay and the Doppler that to be estimated. 
Denote the eigenvalue decomposition (EVD) of $\mathbf{Y}_{u}^{\mathrm{DDS}}\left( \mathbf{Y}_{u}^{\mathrm{DDS}} \right) ^{\mathrm{H}}$ as $\mathbf{U\Lambda U}^{-1}=\mathrm{eig}\left( \mathbf{Y}_{u}^{\mathrm{DDS}}\left( \mathbf{Y}_{u}^{\mathrm{DDS}} \right) ^{\mathrm{H}} \right) $, where $\mathbf{U}$ and $\mathbf{\mathbf{\Lambda }}$ denote the eigenvector matrix and eigenvalue matrix, respectively. Note that $\mathbf{U}$ comprises a complete set of orthogonal bases that span the column space of $\mathbf{Y}_{u}^{\mathrm{DDS}}$. Therefore, there exists a full-rank matrix $\mathbf{M}\in \mathbb{C} ^{P\times P}$ such that 
\begin{equation}\label{estaoa1}
	\begin{aligned}	
	&\mathbf{U}\left( :,1:P \right) \mathbf{M}
	\\
	&\qquad=\left[ \mathbf{a}_{N_{\mathrm{BS}}}\left( \theta _{u,0} \right) ,\mathbf{a}_{N_{\mathrm{BS}}}\left( \theta _{u,1} \right) ,...,\mathbf{a}_{N_{\mathrm{BS}}}\left( \theta _{u,P-1} \right) \right]
	\\
	&\qquad\triangleq {\mathbf{A}}_{N_{\mathrm{BS}},u}.
	\end{aligned}
\end{equation}
The matrix ${\mathbf{A}}_{N_{\mathrm{BS}},u}$ comprises the array steering vectors for the $P$ propagation paths of the $u$-th MS and exhibits a Vandermonde structure. Consequently, we have
\begin{equation}\label{estaoa2}
	\mathbf{U}\left( 2:N_{\mathrm{BS}},1:P \right) \mathbf{M}=\mathbf{U}\left( 1:N_{\mathrm{BS}}-1,1:P \right) \mathbf{M\Theta },
\end{equation}
where 
\begin{equation}\label{similar}
	\mathbf{\Theta }\triangleq \mathrm{D}\left( \left[ e^{j2\pi \frac{d}{\lambda}\cos \theta _{u,0}},e^{j2\pi \frac{d}{\lambda}\cos \theta _{u,1}},...,e^{j2\pi \frac{d}{\lambda}\cos \theta _{u,P-1}} \right] ^{\mathrm{T}} \right).
\end{equation}
Based on (\ref{similar}), we have $(\mathbf{U}(1:N_{\mathrm{BS}}-1, 1:P))^{\dagger} \mathbf{U}(2:N_{\mathrm{BS}},1:P)$ is similar to $\mathbf{\Theta }$, and thus we have
\begin{equation}\label{estaoa3}
	\left( \mathbf{U}\left( 1:N_{\mathrm{BS}}-1,P \right) \right) ^{\dagger}\mathbf{U}\left( 2:N_{\mathrm{BS}},P \right) =\mathbf{M\Theta M}^{-1}.
\end{equation}
Thus, an estimate of $\mathbf{\Theta}$ is obtained by computing the EVD of $\left( \mathbf{U}\left( 1:N_{\mathrm{BS}}-1,P \right) \right) ^{\dagger}\mathbf{U}\left( 2:N_{\mathrm{BS}},P \right)$. Following this, the AoA of each propagation path is subsequently estimated from the eigenvalues of $\mathbf{\Theta}$.
Based on the estimated AOA $\left\{ \hat{\theta}_{u,i} \right\} _{i=0}^{P-1}$, we can reconstruct the matrix ${\hat{\mathbf{A}}}_{N_{\mathrm{BS}},u}$. The channel observation $\left( \hat{\tilde{\mathbf{X}}}_{u}^{\mathrm{DD}}\odot \hat{\tilde{\mathbf{\Phi}}}_{u,i} \right) \hat{\tilde{\mathbf{h}}}_{u,i}^{\mathrm{DD}}$ in the DD domain can be subsequently recovered by projecting the matrix $\mathbf{Y}_{u}^{\mathrm{DDS}}$ onto the subspace orthogonal to the estimated steering array to separate each propagation path in the spatial domain as 
\begin{equation}\label{projection}
	\begin{aligned}
		\hat{\mathbf{Y}}_{u}^{\mathrm{DD}}
		&=\left( \left( \hat{\mathbf{A}}_{N_{\mathrm{BS}},u}^{\mathrm{H}}\hat{\mathbf{A}}_{N_{\mathrm{BS}},u}^{} \right) ^{-1}\hat{\mathbf{A}}_{N_{\mathrm{BS}},u}^{\mathrm{H}}\mathbf{Y}_{u}^{\mathrm{DDS}} \right) ^{\mathrm{T}}
		\\
		&=\left[ \left( {{\mathbf{X}}}_{u}^{\mathrm{DD}}\odot {{\mathbf{\Phi}}}_{u,0} \right) {{\mathbf{h}}}_{u,0}^{\mathrm{DD}},\left( {{\mathbf{X}}}_{u}^{\mathrm{DD}}\odot {{\mathbf{\Phi}}}_{u,1} \right) {{\mathbf{h}}}_{u,1}^{\mathrm{DD}},\right.
		\\
		&\quad...,\left.\left( {{\mathbf{X}}}_{u}^{\mathrm{DD}}\odot {{\mathbf{\Phi}}}_{u,P-1} \right) {{\mathbf{h}}}_{u,P-1}^{\mathrm{DD}}, \right]+\tilde{\mathbf Z}^{\mathrm{DD}}_u,
	\end{aligned}
\end{equation}
where $\tilde{\mathbf Z}^{\mathrm{DD}}_u$ denotes the additive noise and estimation errors for MS $u$.
Next, we estimate the sparse delay-Doppler channel $\left\{ \mathbf{h}_{u,i}^{\mathrm{DD}} \right\} _{i=0}^{P-1}$ based on the observation $\hat{\mathbf{Y}}_{u}^{\mathrm{DD}}$. Without loss of generality, we focus on the observation vector $\mathbf{y}^{\mathrm{DD}}_{u,i}$ for the $i$-th path. The estimation of  $\hat{\mathbf{h}}_{u,i}^{\mathrm{DD}}$ can be formulated as the following compressive sensing problem with a parameter-dependent dynamic grid problem
\begin{equation}\label{linOb1}
\begin{aligned}
	\mathbf{y}_{u,i}^{\mathrm{DD}}&=\left( {\mathbf{X}}_{u}^{\mathrm{DD}}\odot{\mathbf{\Phi}}_{u,i} \right) {\mathbf{h}}_{u,i}^{\mathrm{DD}}+\tilde{\mathbf{z}}_{u,i}^{\mathrm{DD}}\\
	&={\mathbf{X}}_{u}^{\mathrm{DD}}\left( \tau _{u,i},\nu _{u,i} \right) {\mathbf{h}}_{u,i}^{\mathrm{DD}}+\tilde{\mathbf{z}}_{u,i}^{\mathrm{DD}},\\
\end{aligned}
\end{equation}
where $\mathbf{X}_{u}^{\mathrm{DD}}\left( \tau _{u,i},\nu _{u,i} \right) \triangleq\mathbf{X}_{u}^{\mathrm{DD}}\odot \mathbf{\Phi }_{u,i}\in \mathbb{C} ^{M_pN_p\times M_gN_g}$ is the measurement matrix with dynamic grid parameterized by $\tau _{u,i}$ and $\nu _{u,i}$, and $\tilde{\mathbf{z}}_{u,i}^{\mathrm{DD}}$ denotes the combination of additive noise and projection errors.  The accurate measurement matrix $\mathbf{X}_{u}^{\mathrm{DD}}\left( \tau _{u,i},\nu _{u,i} \right)$ is unavailable at first due to the lack of prior knowledge of fractional delay and Doppler. To address this issue, we propose a two-stage off-grid DD domain channel estimation method. In the first stage, according to (\ref{ongridydd}), a coarse estimation of sparse channel matrix $\hat{\mathbf{h}}_{u,i}^{\mathrm{DD}}$ is performed as
\begin{equation}\label{coarse}
\mathbf{y}_{u,i}^{\mathrm{DD}}\approx \left( {\mathbf{X}}_{u}^{\mathrm{DD}}\odot \bar{\mathbf{\Phi}} \right) {\mathbf{h}}_{u,i}^{\mathrm{DD}}+\tilde{\mathbf{z}}_{u,i}^{\mathrm{DD}},
\end{equation}
where the phase compensate matrix $\bar{\mathbf{\Phi}}\left( p,q \right) =e^{j2\pi \frac{k^{\prime}\left( M_{\mathrm{CP}}+l-l^{\prime} \right)}{N\left( M+M_{\mathrm{CP}} \right)}}$ for $p=\left( l-l_{pu} \right) N_p+\left( k-k_p \right) +1$ and $q=l^{\prime}N_g+\left( k^{\prime}+\lceil \frac{N_g}{2} \rceil \right) +1$. We first estimate the coarse sparse channel matrix $\hat{\mathbf{h}}_{u,i}^{\mathrm{DD}}$ by simultaneous-OMP (SOMP) algorithm according to (\ref{coarse}). Based on the coarse sparse channel matrix $\hat{\mathbf{h}}_{u,i}^{\mathrm{DD}}$, we can estimate the delay and Doppler to update the measurement matrix $\mathbf{X}_{u}^{\mathrm{DD}}\left( \tau _{u,i},\nu _{u,i} \right)$. 
The authors of \cite{dingshi} proposed a low-complexity algorithm for fractional Doppler estimation, which relies solely on the magnitude information of $\mathbf{h}_{u,i}^{\mathrm{DD}}$ to estimate the fractional Doppler. In our method, we extend this approach to scenarios with both fractional delay and fractional Doppler. In the support of the $i$-th propagation path, the magnitude of the delay-Doppler channel $h_{u,i}^{\mathrm{DD}}[k^\prime,l^\prime]$ can be expressed as 
\begin{align}\label{Hmag}
		&h_{u,i}^{\mathrm{mag}}\left[ k^\prime ,l^\prime \right] \nonumber
		\\
		&\qquad\triangleq \left| h_{u,i}\left[ k^\prime ,l^\prime \right] \right|
		\nonumber\\
		&\qquad=\left| \alpha _i\Xi _N\left( k^\prime -k_{ u,i}-\kappa _{ u,i} \right) \Xi _M\left( -l^\prime +l_{ u,i}+\iota _{ u,i} \right) \right|.
\end{align}
The magnitude of the function $\Xi _X\left( x \right) $ reaches the peak at $x=0$ and decreases as $x$ moves away from 0 \cite{raviteja}. Hence, the integer Doppler and delay of the $i$-th propagation path can be firstly determined by finding the maximum magnitude $h_{ u,i}^{\mathrm{mag}}$ in $\mathrm{supp}\left( \mathbf{h}_i \right)$
\begin{equation}\label{findint}
	\left\{ \tilde{k}_{ u,i},\tilde{l}_{ u,i}\right\} =\underset{k,l\in \mathrm{supp}\left( \mathbf{h}_i \right)}{\mathrm{arg}\max} h_{ u,i}^{\mathrm{mag}}\left[ k,l \right] .
\end{equation}
Next, we define
\begin{equation}\label{findfrac}
	\begin{aligned}
		\tilde{k}^\prime_{ u,i}=\underset{k\in \left\{ \left[ \tilde{k}_{ u,i}+1 \right] _N,\left[ \tilde{k}_{ u,i}-1 \right] _N \right\}}{\mathrm{arg}\max} h_{ u,i}^{\mathrm{mag}}\left[ k,\tilde{l}_{ u,i} \right] 
		\\
		\tilde{l}^\prime_{ u,i}=\underset{l\in \left\{ \left[ \tilde{l}_{ u,i}+1 \right] _M,\left[ \tilde{l}_{ u,i}-1 \right] _M \right\}}{\mathrm{arg}\max} h_{ u,i}^{\mathrm{mag}}\left[ \tilde{k}_{ u,i},l \right],
	\end{aligned}
\end{equation}
and the Doppler $k_{ u,i}+\kappa _{ u,i}$ and the delay $l_{ u,i}+\iota _{ u,i}$ are between $\tilde{k}_{ u,i}$ and $\tilde{k}^\prime_{ u,i}$, and $\tilde{l}_{ u,i}$ and $\tilde{l}^\prime_{ u,i}$, respectively. Then, according to (\ref{Hmag}), we have the following expressions 
\begin{align}\label{dopplermag}
	\frac{h_{ u,i}^{\mathrm{mag}}\left[ \tilde{k}_{ u,i},\tilde{l}_{ u,i} \right]}{h_{ u,i}^{\mathrm{mag}}\left[ \tilde{k}_{ u,i}^{\prime},\tilde{l}_{ u,i}^{} \right]}&=\frac{\left| \Xi _N\left( -\kappa _{ u,i} \right) \right|}{\left| \Xi _N\left( \tilde{k}_{ u,i}^{\prime}-\tilde{k}_{ u,i}-\kappa _{ u,i} \right) \right|}\nonumber
	\\
	&=\frac{\left| \sin \left( -\frac{\left( \tilde{k}_{ u,i}^{\prime}-\tilde{k}_{ u,i}-\kappa _{ u,i} \right) \pi}{N} \right) \right|}{\left| \sin \left( -\frac{\kappa _{ u,i}\pi}{N} \right) \right|}\nonumber
	\\
	&\approx \frac{\left| \left( \tilde{k}_{ u,i}^{\prime}-\tilde{k}_{ u,i}-\kappa _{ u,i} \right) \right|}{\left| \kappa _{ u,i} \right|},
\end{align}
and 
\begin{align}\label{delaymag}
	\frac{h_{ u,i}^{\mathrm{mag}}\left[ \tilde{k}_{ u,i},\tilde{l}_{ u,i} \right]}{h_{ u,i}^{\mathrm{mag}}\left[ \tilde{k}_{ u,i},\tilde{l}^\prime _{ u,i} \right]}&=\frac{\left| \Xi _M\left( \iota _{ u,i} \right) \right|}{\left| \Xi _M\left( -\left( l_{u, i}-l^\prime _{ u,i} \right) +\iota _{ u,i} \right) \right|}\nonumber
	\\
	& = \frac{\left|\sin \left( -\frac{\left( \tilde{l}^\prime _{ u,i}-\tilde{l}_{ u,i}-\iota _{ u,i} \right) \pi}{M} \right)\right|} {\left|\sin \left( -\frac{\iota _{ u,i}\pi}{M} \right)\right|}\nonumber
	\\
	& \approx \frac{\left| \left( \tilde{l}^\prime_{ u,i}-\tilde{l}_{ u,i}-\iota_{ u,i} \right) \right|}{\left| \iota _{ u,i} \right|}.
\end{align}
Based on (\ref{dopplermag}) and (\ref{delaymag}), we can estimate the fractional part of the Doppler frequency and the fractional part of the delay as
\begin{align}
	\tilde{\kappa}_{u, i}&=\frac{\left( \tilde{k}_{u, i}-\tilde{k}^\prime _{u, i} \right) h_{i}^{\mathrm{mag}}\left[ \tilde{k}^\prime_{u, i},l_{u, i} \right]}{h_{i}^{\mathrm{mag}}\left[ \tilde{k}_{u, i},l_{u, i} \right] +h_{i}^{\mathrm{mag}}\left[ \tilde{k}^\prime _{u, i},l_{u, i} \right]}\label{estdopfrac}\\
	\tilde{\iota}_{u, i}&=\frac{\left( \tilde{l}_{u, i}-\tilde{l}^\prime_{u, i} \right) h_{i}^{\mathrm{mag}}\left[ k_{u, i},\tilde{l}^\prime _{u, i} \right]}{h_{i}^{\mathrm{mag}}\left[ \tilde{k}_{u, i},\tilde{l}_{u, i} \right] +h_{i}^{\mathrm{mag}}\left[ \tilde{k}_{u, i},\tilde{l}^\prime _{u, i} \right]}.\label{estdelfrac}
\end{align}
The channel gain ${\alpha}_i$ is estimated as
\begin{equation}\label{estgain}
	\tilde{\alpha}_{u, i}=\frac{h_{u,i}^{\mathrm{mag}}\left[ \tilde{k}_{u, i},\tilde{l}_{u, i} \right]}{\Xi _N\left( \tilde{k}_{u, i}-k_{u, i}-\kappa _{u, i} \right) \cdot \Xi _M\left( -\tilde{l}_{u, i}+l_{u, i}+\iota _{u,i} \right)}.
\end{equation}
Next, based on the coarse estimate of fractional Doppler and fractional delay, the phase compensation matrix $\tilde{\mathbf{\Phi }}_{u,i}$ is constructed, and the channel parameters are updated based on the updated measurement matrix $\mathbf{X}^{\mathrm{DD}}\left( {\tau }_{u,i},{\nu }_{u,i} \right) =\mathbf{X}^{\mathrm{DD}}\odot \tilde{\mathbf{\Phi}}_{u,i}$. The proposed algorithm is detailed in Algorithm \ref{alg1}.

\begin{figure*} 
	\label{exampleofconvolution}
	\centering
	\subfigure[The DD domain channel for the $u$-th MS at one antenna.]{
		\begin{minipage}{0.31\linewidth}\label{convhDD}
			\includegraphics[width=1.0\linewidth]{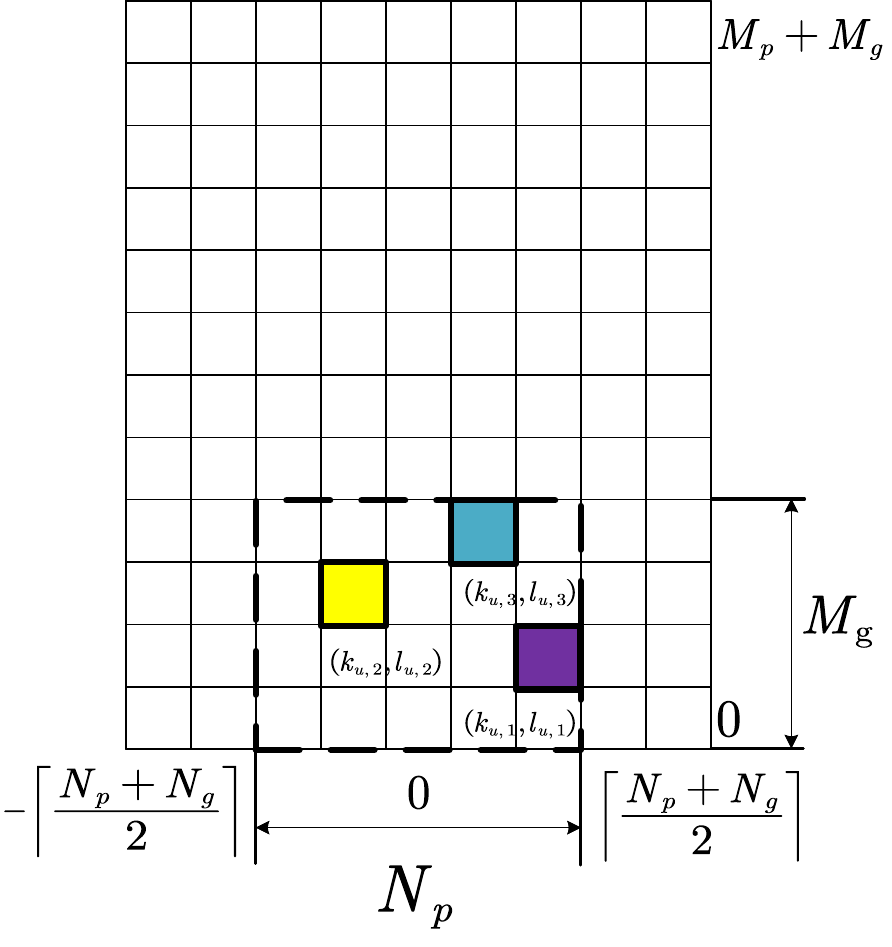} \\
		\end{minipage}
	}
	\subfigure[Pilot Region for MS $u$.]{
		\begin{minipage}{0.32\linewidth}\label{convpilot}
			\includegraphics[width=1.0\linewidth]{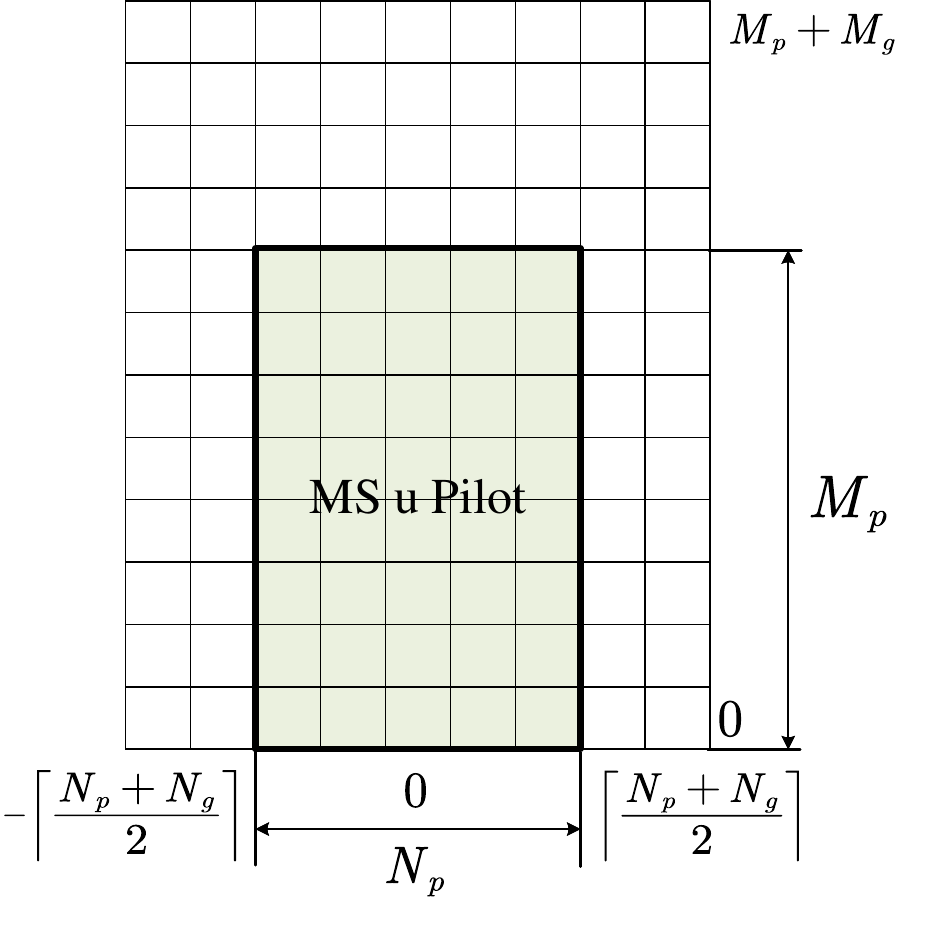} \\
		\end{minipage}
	}
	\subfigure[The received DD domain signal for the $u$-th MS at one antenna.]{
		\begin{minipage}{0.31\linewidth}\label{convyDD}
			\includegraphics[width=1.0\linewidth]{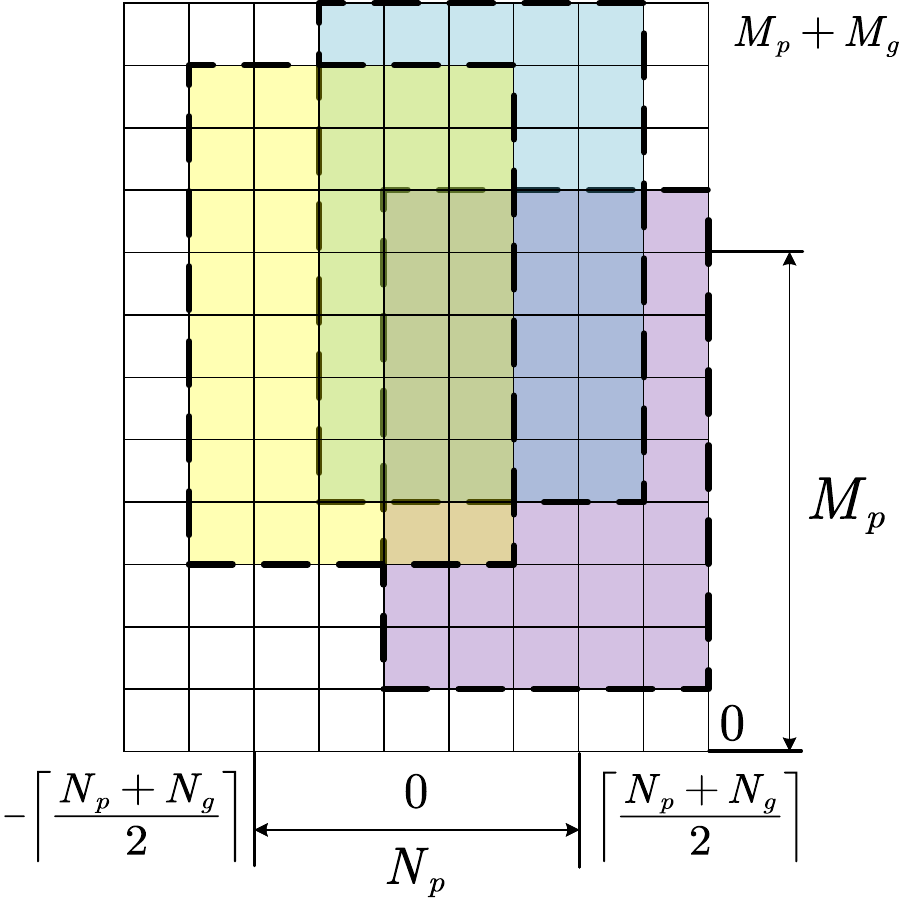} \\
		\end{minipage}
	}
	\caption{The relationship between the received signal $y^{\mathrm{DDS}}\left[ k,l,n_{\mathrm{BS}} \right] $ at one antenna, the pilot and the channel $h^{\mathrm{DDS}}\left[ k,l,n_{\mathrm{BS}} \right] $ at one antenna.}
\end{figure*}

\begin{algorithm}[t]
	\caption{Proposed Channel Estimation Algorithm for Conventional Pilot Structure}
	\label{alg1}
	\begin{algorithmic}[1]
		\REQUIRE $Y^{\mathrm{DDS}}$, $\mathbf{X}^{\mathrm{DD}}$.
		\STATE {\bf {for}} $u=1,2,\ldots ,K$\\
		\STATE Estimate the AoAs $\left\{ \theta _{u,i} \right\} _{i=0}^{P-1}$ by (\ref{estaoa1}), (\ref{estaoa2}) and (\ref{estaoa3}).
		\STATE Calculate the projection by (\ref{projection}).
		\STATE {\bf {for}} $i=0,1,\ldots ,P-1$\\
		\STATE Calculate the coarse estimate $\tilde{\mathbf{h}}^{\mathrm{DD}}_{u,i}$ by SOMP based on $\mathbf{X}^{\mathrm{DD}}$.
		
		\STATE Find the index $\tilde{k}_{u, i},\tilde{l}_{u, i}$, and $\tilde{k}^\prime_{u, i} $ and $\tilde{l}^\prime_{u, i}$ by (\ref{findint}) and (\ref{findfrac}), respectively.
		\STATE Estimate the fractional part Doppler $\tilde{\kappa}_{u, i}$ and the fractional part delay $\tilde{\iota}_{u, i}$ by (\ref{estdopfrac}) and (\ref{delaymag}), respectively.
		\STATE Estimate the channel gain $\tilde{\alpha}_i$ by (\ref{estgain}).
		
		\STATE Construct the phase compensation matrix $\mathbf{\Phi }_{u,i}$ and measurement matrix $\mathbf{X}^{\mathrm{DD}}\left( {\tau }_{u,i},{\nu }_{u,i} \right) $.
		\STATE Update the estimate $\tilde{\mathbf{h}}^{\mathrm{DD}}_{u,i}$ by SOMP based on $\mathbf{X}^{\mathrm{DD}}\left( {\tau }_{u,i},{\nu }_{u,i} \right) $.
		
		\STATE Repeat step 6-8 to obtain the accurate estimate of channel parameters.
				\STATE {\bf {end for}}.
		\STATE {\bf {end for}}.  
		\ENSURE $\left\{ \tilde{k}_{u, i} \right\} , \left\{ \tilde{\kappa}_{u, i} \right\} , \left\{ \tilde{l}_{u, i} \right\} , \left\{ \tilde{\iota}_{u, i} \right\} , \left\{ \tilde{\alpha}_{u, i} \right\} $.  
	\end{algorithmic}
\end{algorithm}

\section{Proposed CSEP Structure and Channel Estimation Method}

In the previous section, we introduced the conventional pilot structure for MU-OTFS and proposed a multi-dimensional decomposition-based channel estimation method.
However, the conventional pilot structure suffers from a significant  overhead of $K\left( M_p+M_g \right) \left( N_p+M_g \right) $, and a considerable portion of which is caused by pilot symbol redundancy. 
To further reduce the pilot overhead, we propose a CSEP structure based on Zadoff-Chu (ZC) sequences. The proposed pilot structure benefits from a reduced total overhead of $\left( KM_p+M_g \right) \left( N_p+M_g \right) $. To demonstrate the feasibility of the proposed pilot scheme, we discuss the correlation of the compressed sensing codewords among different MSs. Finally, we propose a novel channel estimation algorithm based on the proposed CSEP structure.

\subsection{CSEP Structure}
Before introducing the proposed CSEP structure, we first analyze the relationship between the positions of the received signals in the DD domain, the positions of the transmit pilots and the DD domain channel.

Fig.~\ref{convhDD} shows an example of the DDS channel $h_{u}^{\mathrm{DDS}}\left[ k,l,n_{\mathrm{BS}} \right] $ at one antenna of the $u$-th MS, where $h_{u}^{\mathrm{DDS}}\left[ k,l,n_{\mathrm{BS}} \right] \triangleq \sum_{i=0}^{P-1}{}h_{u,i}^{\mathrm{DDS}}\left[ k,l,n_{\mathrm{BS}} \right] $. 
The area within the dashed box represents the support set of the DDS channel at the delay-Doppler plane. The purple, yellow and blue squares denote the non-zero elements within the sparse DDS channel at one antenna, which corresponds to three distinct propagation paths. The coordinate pairs, $\left( k_{u,1},l_{u,1} \right) $, $\left( k_{u,2},l_{u,2} \right) $ and $\left( k_{u,3},l_{u,3} \right) $ indicate the indices of the three paths on the delay-Doppler plane. For convenience, we assume that the delay and Doppler are on-grid in this example. 
Fig.~\ref{convpilot} shows the position of the pilot block for one MS, where the center of the block corresponds to the zero point on the Doppler axis. 
Fig.~\ref{convyDD} presents the DDS domain signal $y^{\mathrm{DDS}}\left[ k,l,n_{\mathrm{BS}} \right] $ at one antenna. The dashed boxes in purple, yellow and blue represent the received signals corresponds to the pilot block in Fig.~\ref{convyDD} after passing through the channel paths corresponding to the purple, yellow, and blue squares in Fig.~\ref{convhDD}, respectively. 
Fig.~\ref{convyDD} demonstrates that the received DDS domain signal $y^{\mathrm{DDS}}\left[ k,l,n_{\mathrm{BS}} \right] $ at one antenna is the result of a 2D convolution between the DDS domain channel $h_{u}^{\mathrm{DDS}}\left[ k,l,n_{\mathrm{BS}} \right] $ at one antenna and the transmitted signal $X_{u}^{\mathrm{DD}}\left[ k,l \right] $. 
Specifically, the received signal $y^{\mathrm{DDS}}\left[ k,l,n_{\mathrm{BS}} \right] $ at one antenna originating from the transmitted signal of the $u$-th MS and propagating through the $i$-th channel path will be shifted in the DD domain by $l_{u,i}$ in the delay dimension and by $k_{u,i}$ in the Doppler dimension relative to the original position of the pilot block $X_{u}^{\mathrm{DD}}\left[ k,l \right] $. 

Therefore, it can be seen that each column in the measurement matrix $\mathbf{X}_{u}^{\mathrm{DD}}$ of the sparse DD domain channel recovery problem (\ref{coarse}) is a \textit{shifted} copy of the vectorized pilot block from Fig.~\ref{convpilot}. If the redundant pilot from Fig.~\ref{individualpilotfig} is employed, then each column in $\mathbf{X}_{u}^{\mathrm{DD}}$ becomes a \textit{cyclically shifted} copy of the vectorized pilot block. This characteristic of the OTFS input-output relationship inspires us to circular shift orthogonality of ZC sequences for multiuser separation.

\begin{figure*}[ht]
	\begin{minipage}[ht]{1\linewidth}
		\centering
		\includegraphics[width=1.0\linewidth]{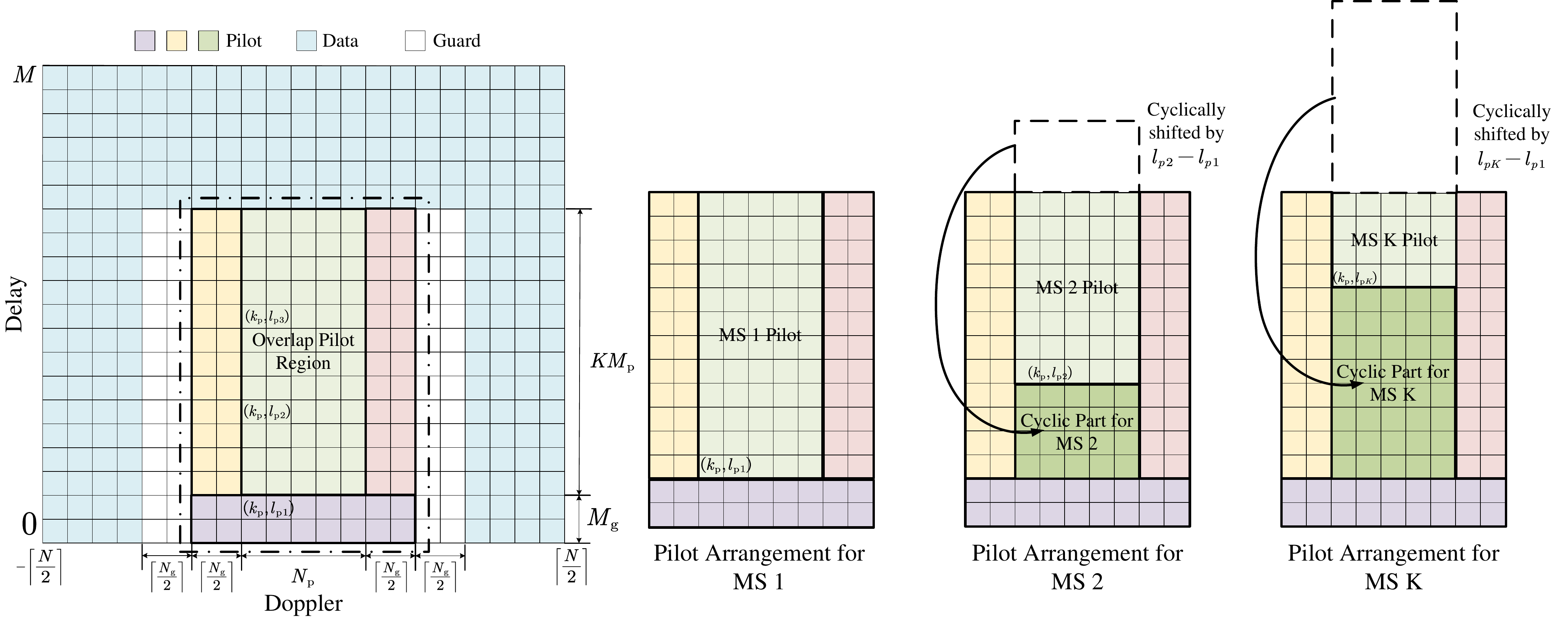}
		\caption{Proposed CSEP structure for MU-OTFS systems.}
		\label{overlappilotfig}
	\end{minipage}%
	\hfill
\end{figure*}

Fig.~\ref{overlappilotfig} illustrates the proposed  CSEP  structure. The dash-dotted boundary marks the common pilot region shared by all MSs. Similar to the scheme in Section \ref{CEmodelandProbForm4IPS}, the green area represents the pilot block for each MS, while the yellow, the red and the purple areas constitute redundant extensions of the green pilot block. However, differing from the conventional pilot structure, each MS's pilot block is of length $KM_{p}$ in delay axis and these pilot blocks overlap within the green region in the proposed  CSEP  structure. For MS $u$, its pilot block  in green region is given by
\begin{equation}\label{Cu}
\mathbf{C}_u=\mathbf{p}_{\left( u-1 \right) M_{p}}^{KM_p}\left( \mathbf{p}_{0}^{N_p} \right) ^T,
\end{equation}  where $\mathbf{p}_{c}^{N}\in \mathbb{C} ^{N\times 1}$ represents a ZC sequence of length $N$ with a cyclic shift of $c$, whose $n$-th element is given by 
\begin{equation}\label{pNc}
p_{c}^{N}\left[ n \right] =\begin{cases}
	\frac{1}{\sqrt{N}}e^{\frac{j\pi \left( \left[ n-c \right] _N \right) ^2}{N}},               &\mathrm{if}\, N\,\,\mathrm{is}\, \mathrm{even}\\
	\frac{1}{\sqrt{N}}e^{\frac{j\pi \left( \left[ n-c \right] _N \right) \left( \left[ n-c+1 \right] _N \right)}{N}}, &\mathrm{if}\, N\,\,\mathrm{is}\, \mathrm{odd}\\
\end{cases}
\end{equation}
where $n=0,1,\ldots,N-1$. From (\ref{Cu}), it can be observed that the pilot symbols for all MSs are identical but undergo different cyclic shifts. Specifically, if the first MS's pilot starts at position $\left( k_p,l_{p1} \right) $, then the pilot block for the $u$-th MS is cyclically shifted by $\left( u-1 \right) M_p$ positions relative to the first MS. This is equivalent to the starting position of the pilot block for MS $u$ being $\left( k_p,l_{pu} \right) $, which is shown in Fig.~\ref{overlappilotfig}, where $l_{pu}=l_{p1}+\left( u-1 \right) M_p$. 

By stacking all the symbols $Y^{\mathrm{DDS}}\left[ k,l,n_{\mathrm{BS}} \right] $ at the $n_{\mathrm{BS}}$-th antenna with indices $k=k_p,k_p+1,...,k_p+N_p-1$ and $l=l_{p1},l_{p1}+1,...,l_{p1}+KM_p-1$, we have 
\begin{equation}\label{yDDSnBSoverlap}
\mathbf{y}_{}^{\mathrm{DD},n_{\mathrm{BS}}}=\sum_u{}\left( \mathbf{X}_{u}^{\mathrm{DD}}\odot {\mathbf{\Phi}}_u \right) \mathbf{h}_{u}^{\mathrm{DD},n_{\mathrm{BS}}}+\mathbf{z}_{}^{\mathrm{DD},n_{\mathrm{BS}}},
\end{equation}
where $\mathbf{X}_{u}^{\mathrm{DD}}\in \mathbb{C} ^{N_pKM_p\times N_gM_g}$ denotes the measurement matrix for MS $u$ with the $l^{\prime}N_g+k^{\prime}+\lceil \frac{N_g}{2} \rceil +1$-th column of $\mathbf{p}_{k^{\prime}}^{N_p}\otimes \mathbf{p}_{\left( u-1 \right) M_p+l^{\prime}}^{KM_p}$, and
\begin{align}
\mathbf{h}_{u}^{\mathrm{DD},n_{\mathrm{BS}}}\left( q \right) &=\begin{cases}
	\left[ \mathbf{a}_{N_{\mathrm{BS}}}\left( \theta _{u,i} \right) \right] _{n_{\mathrm{BS}}}\mathbf{h}^{\mathrm{DD}}_{u,i}\left( q \right) ,q\in \bigcup\nolimits_i^{}{\mathrm{supp}\left( \mathbf{h}_{u,i} \right) \,\,\,\,}\\
	0,\qquad\qquad\qquad\qquad\quad\ \,\, \mathrm{otherwise}\\
\end{cases}
\end{align}
for $p^{\prime}=\left( l-l_{p1} \right) N_p+\left( k-k_p \right) +1$  with $l=l_{p1},l_{p1}+1,...,l_{p1}+KM_p-1, k=k_p,k_p+1,...,k_p+N_p-1$, and  $q=l^{\prime}N_g+\left( k^{\prime}+\lceil \frac{N_g}{2} \rceil \right) +1$ with $k^{\prime}=-\lceil \frac{N_g}{2} \rceil ,-\lceil \frac{N_g}{2} \rceil +1,...,\lceil \frac{N_g}{2} \rceil $ and $l^{\prime}=0,1,...,M_g-1$, respectively.
As can be seen from (\ref{yDDSnBSoverlap}), the multi-MS channel estimation with  CSEP  is still a CS problem. However, the vector $\mathbf{h}^{\mathrm{DD}}$ is the concatenation of channels from all $K$ MSs. In the specific case where the phase compensation matrix $\mathbf{\Phi }$ reduces to an all-one matrix, the channel for the $u$-th MS can be directly estimated via the least square (LS) algorithm, exploiting the shift orthogonality of the ZC sequences as
\begin{equation}
\hat{{\mathbf{h}}}_{u}^{\mathrm{DD},n_{\mathrm{BS}}}=\left( \mathbf{X}_{u}^{\mathrm{DD}} \right) ^{\mathrm{H}}\mathbf{y}_{}^{\mathrm{DD},n_{\mathrm{BS}}}.
\end{equation}  
However, the phase compensation matrix $\mathbf{\Phi }$ depends on the channel parameters to be estimated. Therefore, it is necessary to analyze the column coherence of the measurement matrix $\left( \mathbf{X}^{\mathrm{DD}}\odot \mathbf{\Phi } \right)$.

\begin{algorithm}[t]
	\caption{Proposed Channel Estimation Algorithm for  CSEP  Structure}
	\label{alg2}
	\begin{algorithmic}[1]
		\REQUIRE $\mathbf{Y}^{\mathrm{DDS}}$, $\mathbf{X}^{\mathrm{DD}}_u$.
		\STATE {\bf {for}} $u=1,2,\ldots ,K$\\
		\STATE Find the support $\mathcal{S} _u$ of ${\mathbf{h}}^{\mathrm{DD}}_{u}$ by SOMP based on the measurement matrix $\left( \mathbf{X}_{u}^{\mathrm{DD}}\odot \mathbf{\Phi } \right) $.
		\STATE Calculate $\hat{\mathbf{H}}_{u}^{\mathrm{DDS}}$ by (\ref{calHDDSu}).
		\STATE Calculate the EVD $\mathbf{U\Lambda U}^{\mathrm{H}}=\mathrm{eig}\left( \hat{\mathbf{H}}_{u}^{\mathrm{DDS}}\left( \hat{\mathbf{H}}_{u}^{\mathrm{DDS}} \right) ^{\mathrm{H}} \right) $ and estimate the AoAs by (\ref{estaoa1})-(\ref{estaoa3})
		\STATE Calculate the projection by (\ref{projection}).
		\STATE {\bf {for}} $i=0,1,\ldots ,P-1$\\
		\STATE Find the index $\tilde{k}_{u, i},\tilde{l}_{u, i}$, and $\tilde{k}^\prime_{u, i} $ and $\tilde{l}^\prime_{u, i}$ by (\ref{findint}) and (\ref{findfrac}), respectively.
		\STATE Estimate the fractional part Doppler $\tilde{\kappa}_{u, i}$ and the fractional part delay $\tilde{\iota}_{u, i}$ by (\ref{estdopfrac}) and (\ref{delaymag}), respectively.
		\STATE Estimate the channel gain $\tilde{\alpha}_i$ by (\ref{estgain}).
		\STATE {\bf {end for}}.
		\STATE Construct the phase compensation matrix $\mathbf{\Phi }_{u}$ and the measurement matrix $\mathbf{X}^{\mathrm{DD}}_u\odot \mathbf{\Phi }_u$.
		\STATE Find the support $\mathcal{S} _u$ of ${\mathbf{h}}^{\mathrm{DD}}_{u}$ by SOMP based on the measurement matrix $\left( \mathbf{X}_{u}^{\mathrm{DD}}\odot \mathbf{\Phi }_u \right) $.
		\STATE Repeat step 5-10 to obtain the accurate estimate of channel parameters.
		\STATE {\bf {end for}}.  
		\ENSURE $\left\{ \tilde{k}_{u, i} \right\} , \left\{ \tilde{\kappa}_{u, i} \right\} , \left\{ \tilde{l}_{u, i} \right\} , \left\{ \tilde{\iota}_{u, i} \right\} , \left\{ \tilde{\alpha}_{u, i} \right\} $.

	\end{algorithmic}
\end{algorithm}

\subsection{Coherence Analysis}
In this subsection, we discuss the coherence between the columns corresponding to different MSs in the measurement matrix $\mathbf{X}^{\mathrm{DD}}\odot \mathbf{\Phi }$. Specifically, consider MS $s$ and MS $t$ $(s\neq t)$. The $i$-th path of MS $s$ has delay $l_{s,i}+\iota _{s,i}$ and Doppler $k_{s,i}+\kappa _{s,i}$, and it exhibits a peak in  $\mathbf{h}_{s}^{\mathrm{DD},n_{\mathrm{BS}}}$ at $l_{s,i}N_g+k_{s,i}+\lceil \frac{N_g}{2} \rceil +1$. The $j$-th path of MS $t$ has delay $l_{t,j}+\iota _{t,j}$ and Doppler $k_{t,j}+\kappa _{t,j}$, and it exhibits a peak in  $\mathbf{h}_{t}^{\mathrm{DD},n_{\mathrm{BS}}}$ at the $l_{t,j}N_g+k_{t,j}+\lceil \frac{N_g}{2} \rceil +1$.
With knowledge of the delay and Doppler, we denote the column associated with the peak of the $i$-th path of MS $s$ in $\mathbf{h}_{s}^{\mathrm{DD},n_{\mathrm{BS}}}$ is $\mathbf{c}_{s,i}^{\mathrm{peak}}$, which is given by
\begin{equation}\label{c_si}
	\begin{aligned}
		\mathbf{c}_{s,i}^{\mathrm{peak}}&=\mathbf{X}^{\mathrm{DD}}_u\left( :,l_{s,i}N_g+k_{s,i}+\lceil \frac{N_g}{2} \rceil +1 \right) 
		\\
		&\quad\odot \mathbf{\Phi }_u\left( :,l_{s,i}N_g+k_{s,i}+\lceil \frac{N_g}{2} \rceil +1 \right) 
		\\
		&=\mathbf{p}_{k_{s,i}}^{N_p}\otimes \mathbf{p}_{\left( s-1 \right) M_p+l_{s,i}}^{KM_p}
		\\
		&\quad\odot \mathbf{\Phi }_u\left( :,l_{s,i}N_g+k_{s,i}+\lceil \frac{N_g}{2} \rceil +1 \right), 
	\end{aligned}
\end{equation}
where the $\left( l-l_{p1} \right) N_p+\left( k-k_p \right) +1$-th element in vector $\mathbf{\Phi }_u\left( :,l_{s,i}N_g+k_{s,i}+\lceil \frac{N_g}{2} \rceil +1 \right) $ is $e^{j2\pi \frac{\left( k_{s,i}+\kappa _{s,i} \right) \left( M_{\mathrm{CP}}+l-l_{s,i}-\iota _{s,i} \right)}{N\left( M+M_{\mathrm{CP}} \right)}}$, $k=k_p,k_p+1,...,k_p+N_p-1$ and $l=l_{p1},l_{p1}+1,...,l_{p1}+KM_p-1$.
Similarly, we denote the column associated with the peak of the $j$-th path of MS $t$ in $\mathbf{h}_{t}^{\mathrm{DD},n_{\mathrm{BS}}}$ is $\mathbf{c}_{t,j}^{\mathrm{peak}}$, which is given by
\begin{equation}\label{c_tj}
	\begin{aligned}
\mathbf{c}_{t,j}^{\mathrm{peak}}&=\mathbf{X}^{\mathrm{DD}}_t\left( :,l_{t,j}N_g+k_{t,j}+\lceil \frac{N_g}{2} \rceil +1 \right) 
\\
&\quad\odot \mathbf{\Phi }_t\left( :,l_{t,j}N_g+k_{t,j}+\lceil \frac{N_g}{2} \rceil +1 \right) 
\\
&=\mathbf{p}_{k_{t,j}}^{N_p}\otimes \mathbf{p}_{\left(t-1\right)M_p+l_{t,j}}^{KM_p}
\\
&\quad\odot \mathbf{\Phi }_t\left( :,l_{t,j}N_g+k_{t,j}+\lceil \frac{N_g}{2} \rceil +1 \right) ,
	\end{aligned}
\end{equation}
where the $\left( l-l_{p1} \right) N_p+\left( k-k_p \right) +1$-th element in vector $\mathbf{\Phi }\left( :,l_{t,j}N_g+k_{t,j}+\lceil \frac{N_g}{2} \rceil +1 \right) $ is $e^{j2\pi \frac{\left( k_{t,j}+\kappa _{t,j} \right) \left( M_{\mathrm{CP}}+l-l_{t,j}-\iota _{t,j} \right)}{N\left( M+M_{\mathrm{CP}} \right)}}$. Denote the coherence between $\mathbf{c}_{s,i}^{\mathrm{peak}}$ and $\mathbf{c}_{t,j}^{\mathrm{peak}}$ by $\mu _{s,i; t,j}=\left| \left( \mathbf{c}_{t,j}^{\mathrm{peak}} \right) ^{\mathrm{H}}\mathbf{c}_{s,i}^{\mathrm{peak}} \right|$, we have the following theorem.
\begin{Theorem}\label{theorem2}
	(Bounded Column Coherence for  CSEP  Strcuture)
	
	Let $\mathbf{c}_{s,i}^{\mathrm{peak}}$ and $\mathbf{c}_{t,j}^{\mathrm{peak}}$ be the columns of the measurement matrix $\mathbf{X}^{\mathrm{DD}}\odot \mathbf{\Phi }$ corresponding to the peak locations of the $i$-th path of MS $s$ and the $j$-th path of MS $t$ in the delay Doppler domain, respectively. The mutual coherence $\mu _{s,i; t,j}=\left| \left( \mathbf{c}_{t,j}^{\mathrm{peak}} \right) ^{\mathrm{H}}\mathbf{c}_{s,i}^{\mathrm{peak}} \right|$ attains its maximum when $k_{s,i}+\kappa _{s,i}-k_{t,j}-\kappa _{t,j}=N_g$ and $M_p=M_g+1$. And its maximum can be upper-bounded by $\epsilon $ if $KM_p< \frac{\epsilon N\left( M+M_{\mathrm{CP}} \right)}{N_g}$.
\end{Theorem}

\textit{Proof:} The proof is provided in Appendix \ref{proofoftheorem2}. \hfill $\blacksquare$

Theorem \ref{theorem2} analyzes an extreme case where $M_p=M_g+1$, deriving the condition that $KM_p$ must satisfy to ensure the maximum coherence remains below a threshold $\epsilon$. In this case, the maximum coherence  $\mu _{s,i;t,j}^{\max}$ occurs near the first zero point of the sinc function in (\ref{usitj}). Due to the property of the sinc function, the maximum magnitude of sidelobes decrease as they move far from the main lobe. Therefore, when $M_p$ increases, the maximum correlation is consequently reduced. 
\begin{figure}[ht]
	\begin{minipage}[ht]{1\linewidth}
		\centering
		\includegraphics[width=1.0\linewidth]{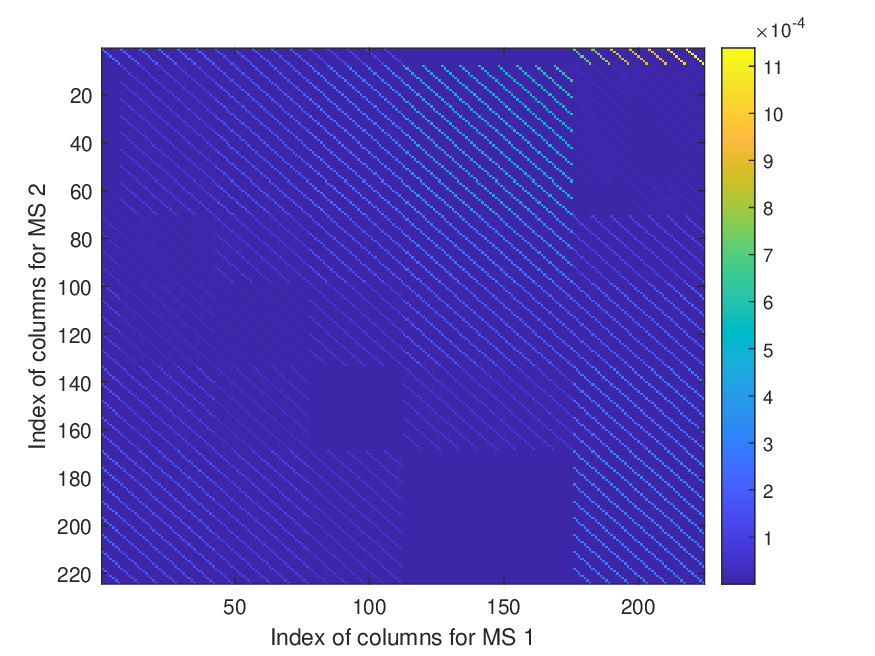}
		\caption{An example of the column-wise coherence between the measurement matrices of MS 1 and MS 2.}
		\label{overlapCoherence}
	\end{minipage}%
	\hfill
\end{figure}
Fig.~\ref{overlapCoherence} shows the column-wise coherence between the $\mathbf{X}^{\mathrm{DD}}\odot \mathbf{\Phi }$ columns corresponding to MS 1 and MS 2, whose pilot blocks have an initial position offset of $M_p = 1.15M_g$. It can be observed that their maximum coherence is on the order of $10^{-3}$, which can be considered nearly orthogonal.

\begin{table*}[t]
	\caption{Computational Complexity of Different Algorithms}
	\label{complexityTable}
	\centering
	\renewcommand{\arraystretch}{1.5} 
	
	\begin{tabular}{>{\centering\arraybackslash}m{3.5cm} >{\centering\arraybackslash}m{13.5cm}}
		\hline
		\textbf{Algorithms} & \textbf{Computational Complexity} \\
		\hline
		\makecell{Proposed Algorithm for \\ Conventional Pilot Structure} & \raisebox{-.3ex}{$\mathcal{O} \left( N_{\mathrm{BS}}^{2}N_gM_g+N_{\mathrm{BS}}^{3}+N_{\mathrm{BS}}^{2}N_pM_pN_gM_g+N_{\mathrm{BS}}^{3}+PN_pM_pN_gM_g+PN_gM_g \right) $} \\
		\hline
		\vspace{5pt}\makecell{Proposed Algorithm  \\ for CSEP Structure}\vspace{5pt} & \raisebox{+.2ex}{$\mathcal{O} \left( PKM_pN_pM_gN_g+N_{\mathrm{BS}}KM_pN_pM_gN_g+N_{\mathrm{BS}}^{2}N_gM_g+N_{\mathrm{BS}}^{3}+PN_{\mathrm{BS}}N_gM_g+PM_gN_g \right) $} \\
		\hline
		\makecell{SBL-based method \\ \cite{sblotfs,shan,VBI,2dsbl}} & \raisebox{-.2ex}{$\mathcal{O} \left(N_{\mathrm{iter}}\left( N_{\mathrm{BS}}^{2}M_pN_p+N_{\mathrm{BS}}^{3}+\left( M_gN_g \right) ^3+\left( M_gN_g \right) ^2M_pN_p+M_gN_gM_pN_p \right)\right) $} \\
		\hline
		\makecell{3D-SOMP method \\ \cite{dingshi,3dsomp}} & \raisebox{-.2ex}{$\mathcal{O} \left( PN^2_{\mathrm{BS}}M_pN_pM_gN_g+PN_{\mathrm{BS}}M_gN_g+GN_{\mathrm{BS}} \right) $} \\
		\hline
	\end{tabular}
	\label{tab5}
\end{table*}

\subsection{Proposed Channel Estimation Method for CSEP Structure}
In this subsection, we illustrate the proposed channel estimation method for the CSEP structure. Stacking all the signals at each antenna, we have
\begin{equation}
\begin{aligned}
	\mathbf{y}_{}^{\mathrm{DDS}} &=\left[\left(\mathbf{Y}_{}^{\mathrm{DD},1}\right)^{\mathrm T}, \left( \mathbf{y}_{}^{\mathrm{DD},2} \right) ^{\mathrm{T}},...,\left( \mathbf{y}_{}^{\mathrm{DD},N_{\mathrm{BS}}} \right) ^{\mathrm{T}} \right] ^{\mathrm{T}}
	\\
	&=\sum_u{}\mathbf{H}_{u}^{\mathrm{DDS}}\left( \mathbf{X}_{u}^{\mathrm{DD}}\odot \mathbf{\Phi }_u \right) ^{\mathrm{T}}+\mathbf{Z}^{\mathrm{DDS}},
\end{aligned}
\end{equation}
where $\mathbf{H}_{u}^{\mathrm{DDS}}$ is given in (\ref{H^DDS_u}) and $\mathbf{Z}^{\mathrm{DDS}}$ is the additive noise. 
The  $\mathbf{Y}_{}^{\mathrm{DDS}}$ is the composite result of the overlap DDS domain signals from all MSs. Consequently, we first separate the signal of the MS to be estimated by leveraging the low coherence between the pilot blocks assigned to different MSs. 
Consider the received signal on the first antenna 
\begin{equation}
\mathbf{y}_{}^{\mathrm{DD},1}=\sum_u{}\left( \mathbf{X}_{u}^{\mathrm{DD}}\odot \mathbf{\Phi }_u \right) \mathbf{h}_{u}^{\mathrm{DD},1}+\mathbf{z}_{}^{\mathrm{DD},1}.
\end{equation}
For MS $u$, we first employ the measurement matrix $\mathbf{X}_{u}^{\mathrm{DD}}\odot \mathbf{\Phi }$ with the SOMP algorithm to find the atomic indices $\mathcal{S} _u$ of the support set for the DD domain channel $\mathbf{h}_{u}^{\mathrm{DD},1}$, where $\bar{\mathbf{\Phi}}$ is given in (\ref{coarse}). Subsequently, the DDS channel of MS $u$ is obtained by
\begin{equation}\label{calHDDSu}
	\begin{aligned}
\hat{\mathbf{H}}_{u}^{\mathrm{DDS}}&=\mathbf{Y}^{\mathrm{DDS}}\left( \left[ \left( \mathbf{X}_{u}^{\mathrm{DD}}\odot \mathbf{\Phi } \right) \right] _{:,\mathcal{S} _u}^{\dagger} \right) ^{\mathrm{T}}.
	\end{aligned}
\end{equation}
Similar to Section \ref{CEforIndividual}, the AoAs of MS $u$ can be estimated via the subspace-based method. Denote the EVD of the correlation matrix $\hat{\mathbf{H}}_{u}^{\mathrm{DDS}}\left( \hat{\mathbf{H}}_{u}^{\mathrm{DDS}} \right) ^{\mathrm{H}}$ as $\mathbf{U\Lambda U}^{\mathrm{H}}=\mathrm{eig}\left( \hat{\mathbf{H}}_{u}^{\mathrm{DDS}}\left( \hat{\mathbf{H}}_{u}^{\mathrm{DDS}} \right) ^{\mathrm{H}} \right) $, the AoAs of MS $u$ can be estimated by (\ref{estaoa1})-(\ref{estaoa3}).
Then, the projection can be computed as
\begin{align}\label{projectionoverlap}
	&\left[ \hat{\mathbf{h}}_{u,0}^{\mathrm{DD}},\hat{\mathbf{h}}_{u,1}^{\mathrm{DD}},...,\hat{\mathbf{h}}_{u,P-1}^{\mathrm{DD}} \right]\nonumber \\
	&\quad=\left( \left( \hat{\mathbf{A}}_{N_{\mathrm{BS}},u}^{\mathrm{H}}\hat{\mathbf{A}}_{N_{\mathrm{BS}},u}^{} \right) ^{-1}\hat{\mathbf{A}}_{N_{\mathrm{BS}},u}^{\mathrm{H}}\hat{\mathbf{H}}_{u}^{\mathrm{DDS}} \right) ^{\mathrm{T}}.
\end{align}
The delay and Doppler are then estimated based on the coarse estimate $\left[ \hat{\mathbf{h}}_{u,0}^{\mathrm{DD}},\hat{\mathbf{h}}_{u,1}^{\mathrm{DD}},...,\hat{\mathbf{h}}_{u,P-1}^{\mathrm{DD}} \right]$ to update the measurement matrix $\mathbf{X}_{u}^{\mathrm{DD}}\odot \mathbf{\Phi }_u$.
The accuracy estimate of delay and Doppler are updated based on the updated measurement matrix. The proposed algorithm for the CSEP structure is detailed in Algorithm \ref{alg2}.

\begin{table}[ht]
	\centering
	\caption{Simulation Parameters}
	\label{simparameters}
	\begin{tabular}{|c|c|}
		\hline
		\textbf{Parameter} & \textbf{Values} \\ 
		\hline
		Carrier frequency (GHz) & 15 \\
		\hline
		Subcarrier spacing (kHz) & 60 \\
		\hline
		Size of OTFS frame \((M, N)\) & (1024,31) \\
		\hline
		Length of CP \(M_{\text{CP}}\) & 72 \\
		\hline
		Guard interval  \((M_g, N_g)\) & (72,7)	\\
		\hline
		Pilot length  \((M_p, N_p)\) & (83,7)	\\
		\hline
		
		Number of BS antennas & 16 \\
		\hline
		Number of MS antennas & 1 \\
		\hline
		Number of MSs & 4 \\
		\hline
		The number of propagation paths & 5 \\
		\hline
		Maximum MS velocity (km/h) & 300 \\
		\hline
		Symbol Modulation Scheme & 16 QAM \\
		\hline
	\end{tabular}
\end{table}

\begin{figure}[t]
	\begin{minipage}[ht]{1\linewidth}
		\centering
		\includegraphics[width=1.0\linewidth]{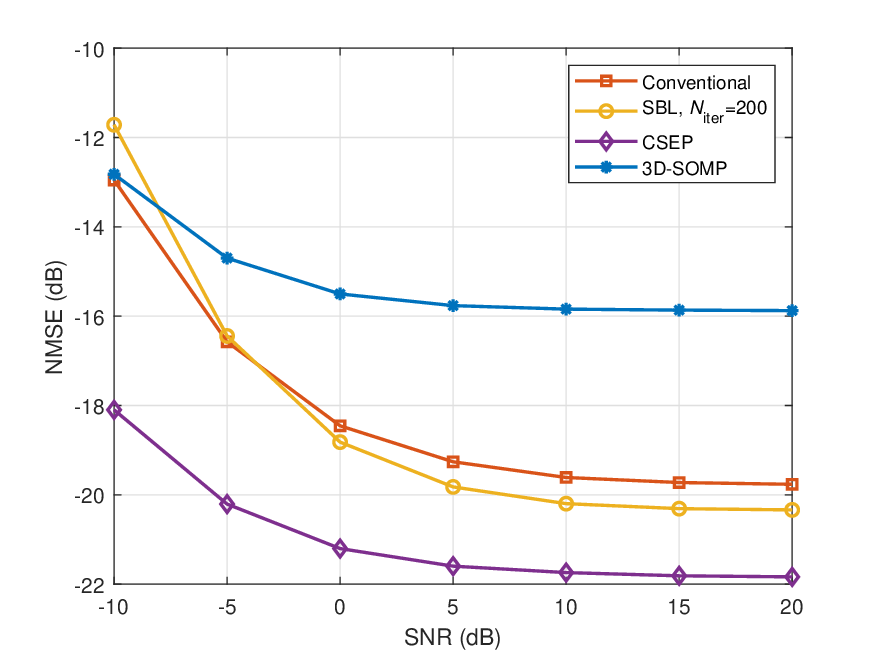}
		\caption{The NMSE performance versus SNR. The number of MSs is 4.}
		\label{NMSEvsSNR}
	\end{minipage}%
	\hfill
\end{figure}

\begin{figure}
		\begin{minipage}[htbp]{1.0\linewidth}
		\centering
		\includegraphics[width=1.0\linewidth]{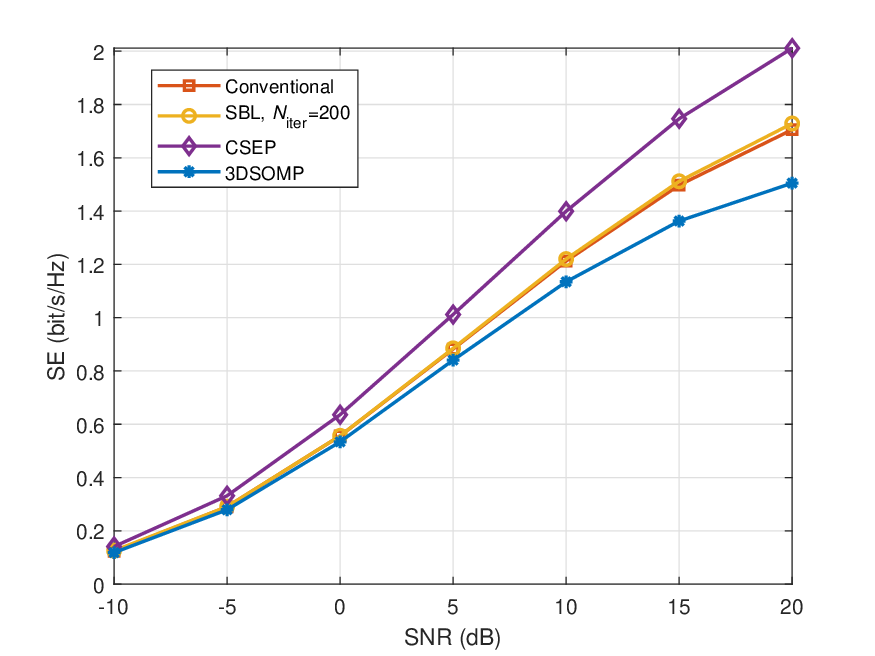}
		\caption{The SE performance versus SNR. The number of MSs is 6.}
		\label{RatevsSNR}
	\end{minipage}%

\end{figure}

\section{Complexity Analysis}\label{CRB}

In this section, the computational complexity of the proposed methods is analyzed. The main complexity of the two proposed algorithms lies in the matrix multiplication of the SOMP algorithm.
Table \ref{tab5} demonstrates the computational complexity of different algorithms for MIMO-OTFS multiuser channel estimation. 
It can be observed that the SOMP matrix multiplication complexity of the proposed method for the CSEP structure is $K$ times higher than that of the proposed method for the conventional pilot structure. 
The complexity of the other steps in both algorithms is comparable. 
Simulation results demonstrate that the proposed CSEP structure achieves an effective trade-off among computational complexity, NMSE, and pilot overhead.
Furthermore, the computational complexity of the SBL-based method is primarily dominated by matrix inversion, with a complexity order of $\mathcal{O}\left( N_{\mathrm{iter}}\left(M_gN_g \right) ^3\right)$, where $N_{\mathrm{iter}}$ denotes the number of iterations. 
The complexity of the 3D-SOMP algorithm is mainly attributed to matrix multiplication involving a measurement of size $N_{\mathrm{BS}}M_pN_p\times N_{\mathrm{BS}}M_gN_g$, resulting in a complexity of $\mathcal{O} \left( PN^2_{\mathrm{BS}}M_pN_pM_gN_g\right)$.

\section{Simulation Results}\label{simulation}
In this section, we present simulation results to show the performance of the proposed two channel estimation methods and the comparison with benchmarks. The simulation parameters are given in Table \ref{simparameters}. The signal-to-noise ratio (SNR) is defined as $\mathrm{SNR}\triangleq \frac{P_s\sigma _{\mathrm{PL}}^{2}}{\sigma _{\mathrm{n}}^{2}}$, where $P_s$, $\sigma _{\mathrm{PL}}^{2}$ and $\sigma _{\mathrm{n}}^{2}$ denote the average symbol power, the pathloss and the noise power, respectively. In the simulation, the pilot symbols and data symbols have the same average power. The pilot length is set to $M_p=1.15M_g=$  83.
The MS velocities are set to 300 km/h, resulting a maximum Doppler extension of $\nu _{\max}=$ 2.304.

\begin{figure}
	\begin{minipage}[htbp]{1.0\linewidth}
	\centering
	\includegraphics[width=1.0\linewidth]{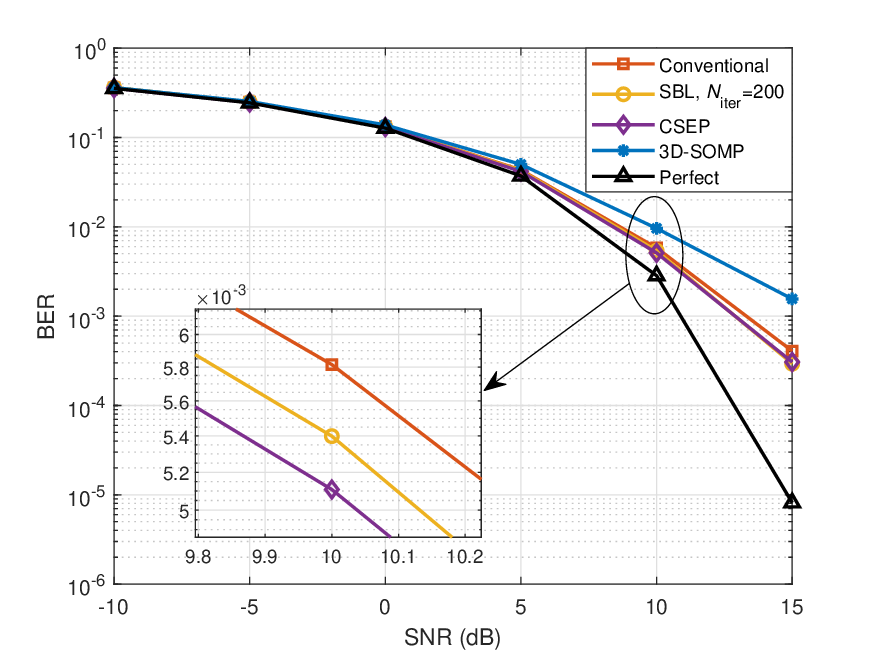}
	\caption{The BER performance versus SNR. The number of MSs is 4.}
	\label{BERvsSNR}
	\end{minipage}%
	\hfill
\end{figure}

\begin{figure}[t]

	\begin{minipage}[htbp]{1.0\linewidth}
	\centering
	\includegraphics[width=1.0\linewidth]{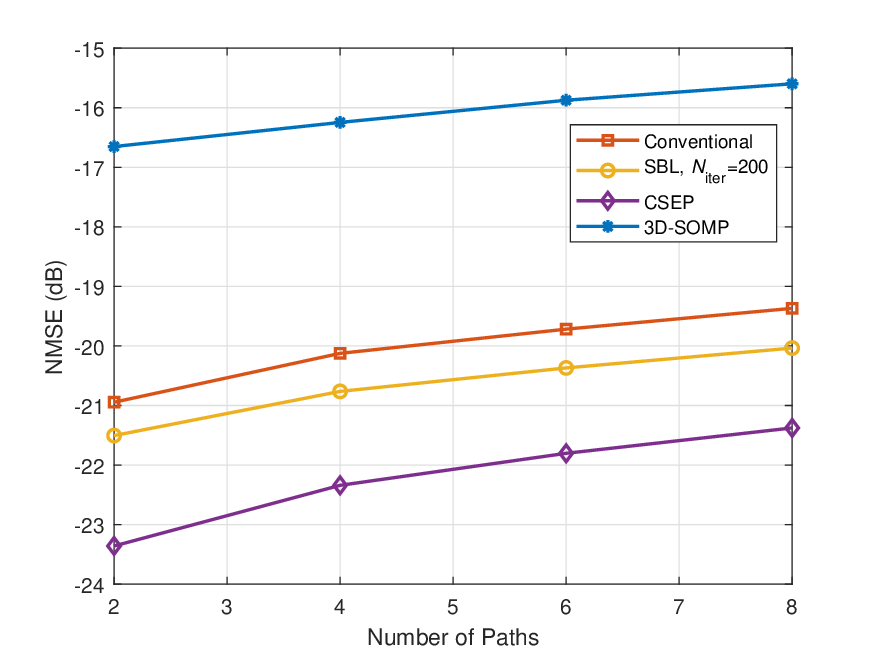}
	\caption{The NMSE performance versus the number of paths. The number of MSs is 4. The SNR is 15 dB.}
	\label{NMSEvsPathNum}
\end{minipage}%
\hfill
\end{figure}

\begin{figure}
	\begin{minipage}[htbp]{1.0\linewidth}
	\centering
	\includegraphics[width=1.0\linewidth]{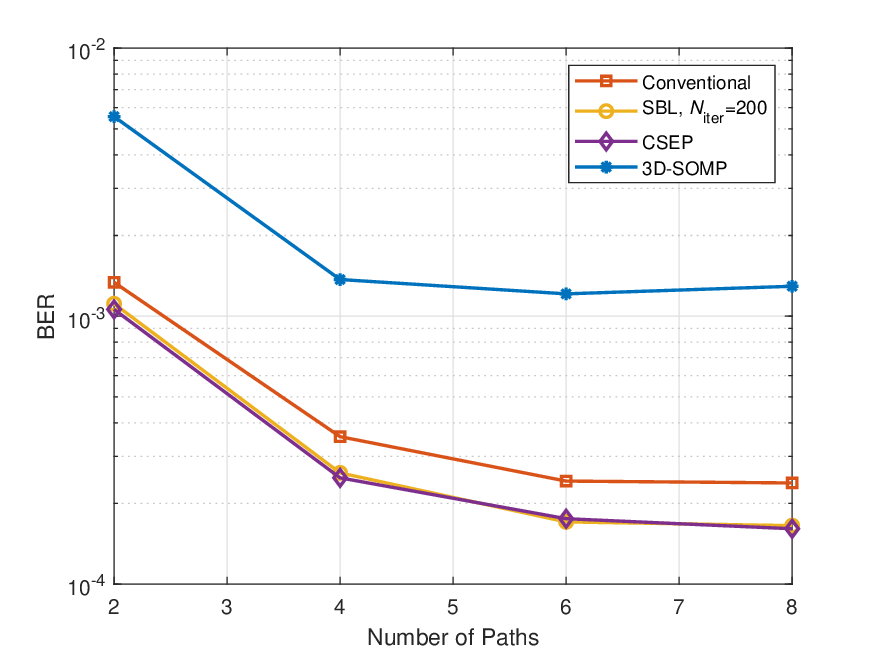}
	\caption{The NMSE performance versus the BER performance. The number of MSs is 4. The SNR is 15 dB.}
	\label{BERvsPathNum}
\end{minipage}%
\hfill
\end{figure}

\begin{figure}
	\begin{minipage}[htbp]{1.0\linewidth}
	\centering
	\includegraphics[width=1.0\linewidth]{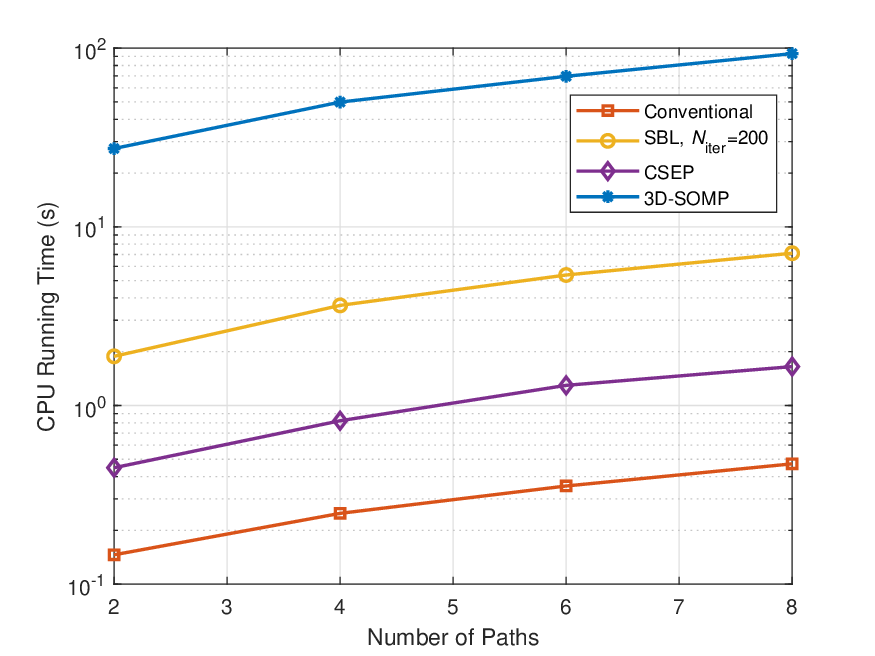}
	\caption{The NMSE performance versus the number of paths. The number of MSs is 4. The SNR is 15 dB.}
	\label{CPUTimevsPathNum}
\end{minipage}%
	\hfill
\end{figure}

\subsection{Performance Comparison with Existing Methods}
We focus on the normalized mean square error (NMSE) performance of the channel estimation. The SBL-based method \cite{sblotfs,shan,VBI,2dsbl} and the 3-D SOMP algorithm \cite{dingshi,3dsomp} are selected as benchmarks, which are denoted as ``SBL-based", ``3D-SOMP", respectively. The above benchmarks use the conventional pilot structure. The proposed channel estimation algorithm for conventional pilot structure and the proposed channel estimation algorithm for the CSEP structure are denoted as ``Conventional", ``CSEP", respectively. The NMSE is defined as
\begin{equation}
	\mathrm{NMSE}\triangleq \mathbb{E} \left\{ \frac{1}{K}\sum_{u=1}^K{\frac{\left\| \hat{\mathbf{H}}_{u}^{\mathrm{DDS}}-\mathbf{H}_{u}^{\mathrm{DDS}} \right\| _{F}^{2}}{\left\| \mathbf{H}_{u}^{\mathrm{DDS}} \right\| _{F}^{2}}} \right\}.
\end{equation}

Fig.~\ref{NMSEvsSNR} shows the NMSE performance versus the SNR for the proposed methods and benchmarks. From Fig.~\ref{NMSEvsSNR}, the proposed CSEP structure and channel estimation method achieve superior NMSE performance compared to other methods.
As shown in (\ref{findint}) and (\ref{findfrac}), the proposed methods utilizes only the main lobe and the strongest side lobe of each path in the DD plane. Although the CSEP structure introduces interference from other MSs, the higher amplitudes of the main lobe and strongest side lobe compared to other side lobes make them more resilient to multi-user interference, thereby minimizing its impact on channel estimation.
Furthermore, the proposed algorithm enables longer pilot sequences and exhibits enhanced noise resistance. Consequently, the CSEP structure and the proposed channel estimation method achieves superior NMSE performance compared to the benchmarks.

In Fig.~\ref{RatevsSNR}, the SE performance versus SNR is demonstrated, where the number of MSs is 6, and the other parameters remain consistent with Table \ref{simparameters}. The SE is calculated as follows:
\begin{equation}
	\mathrm{SE}=\left( 1-\eta \right) \log _2\left( 1+\mathrm{SINR} \right),
\end{equation}
where $\eta$ denotes the pilot overhead in the DD domain, $\eta =\eta ^{\mathrm{i}}$ for individual pilot structure and $\eta =\eta ^{\mathrm{o}}$ for the CSEP structure, and $\eta ^{\mathrm{i}}$ and $\eta ^{\mathrm{o}}$ are given by
\begin{align}
\eta ^{\mathrm{i}}&=\left( K\left( M_p+M_g \right) +M_g \right) \left( N_p+N_g \right)  /\left( NM \right) 	\\
\eta ^{\mathrm{o}}&=\left( KM_p+2M_g \right) \left( N_p+N_g \right)/\left( NM \right). 
\end{align}
When the number of MSs is 6, the pilot overhead for the individual pilot structure is 0.3052, while that for the CSEP structure is only 0.2099, representing a significant reduction of approximately 32\%.  As shown in Fig.~\ref{RatevsSNR}, the CSEP structure achieves a SE gain of 1.1 to 1.2 times compared to the individual pilot structure, which improves the spectrum utilization efficiently.

\begin{figure*}[htbp] 
	\centering
	\subfigure[The BER performance versus SNR.]{
		\begin{minipage}{0.48\linewidth}\label{OTFSvsOFDM_BER}
			\includegraphics[width=1.0\linewidth]{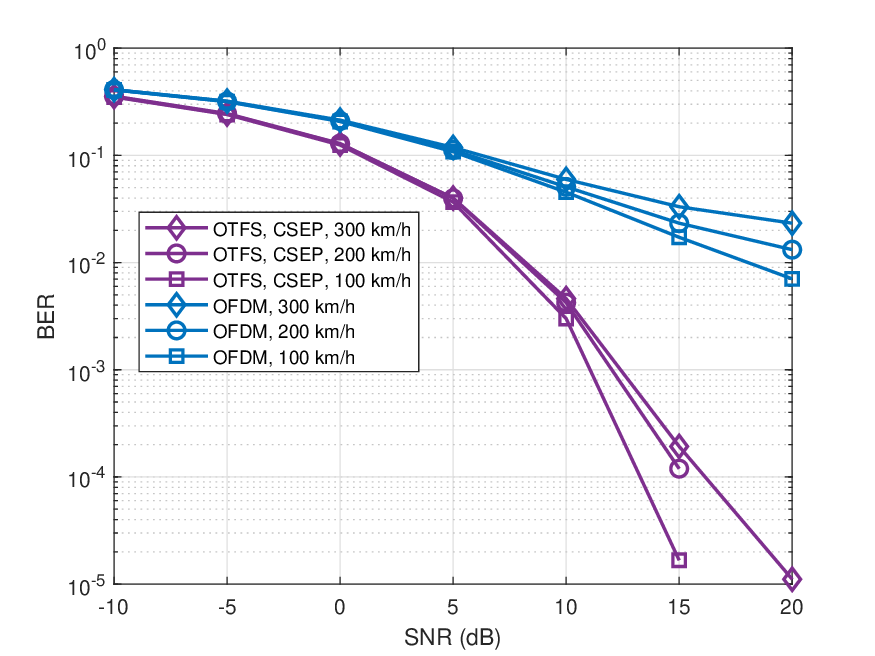} \\
		\end{minipage}
	}
	\subfigure[The SE performance versus SNR.]{
		\begin{minipage}{0.48\linewidth}\label{OTFSvsOFDM_SE}
			\includegraphics[width=1.0\linewidth]{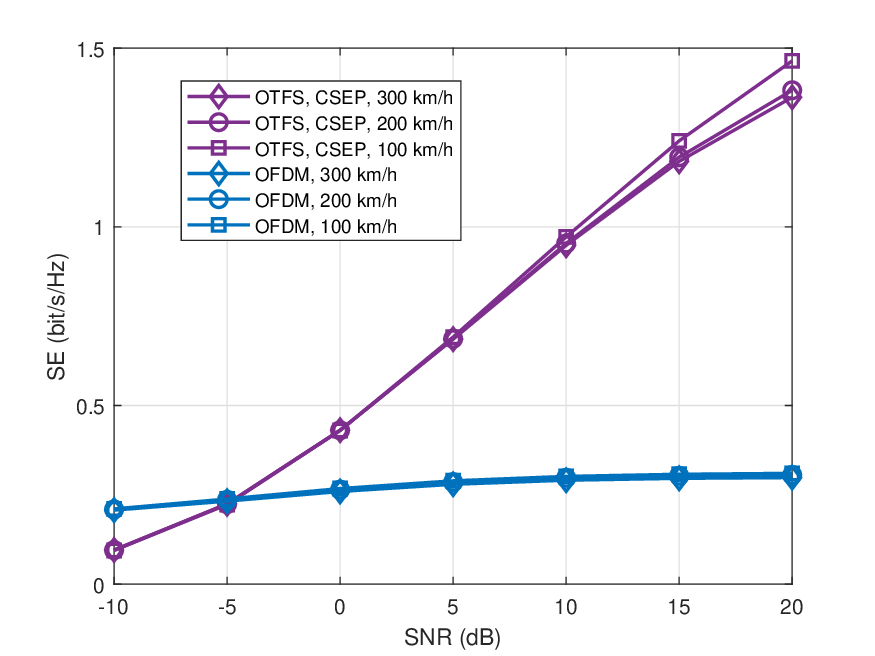} \\
		\end{minipage}
	}
	\caption{Performance comparison of OTFS and OFDM for different maximum MS velocities.}
	\label{OTFSvsOFDM}
\end{figure*}
The BER performance versus the SNR is shown in Fig.~\ref{BERvsSNR}. The detection method is the least square (LS). The transmitted symbols in the DD domain are modulated by 16-QAM. The channel and system parameters are consistent with those specified in Table \ref{simparameters}. Furthermore, the BER performance under perfect CSI is also provided, serving as a performance upper bound. As observed from Fig.~\ref{BERvsSNR}, the BER performance of the CSEP structure approaches that of the perfect CSI across the -10 dB to 10 dB SNR range, and outperforms the individual pilot structure.

The BER performance versus the number of propagation paths are shown in Fig.~\ref{BERvsPathNum}. As illustrated in Fig.~\ref{NMSEvsPathNum} and Fig.~\ref{BERvsPathNum}, while the NMSE performance of all channel estimation algorithms exhibits a corresponding degradation with the increase in the number of propagation paths, the BER performance demonstrates a contrasting upward trend. Such phenomenon can be attributed to the high diversity gain characteristic inherent in OTFS modulation under finite SNR conditions \cite{8686339}. Furthermore, it can be observed that the proposed algorithm maintains satisfactory robustness with the increase of propagation paths. Notably, the BER performance of the proposed algorithm with the CSEP structure demonstrates significant superiority over other benchmark schemes.

Fig.~\ref{NMSEvsPathNum} shows the NMSE performance versus the number of propagation paths. 
For ​the fractional delay and Doppler scenarios, the sparsity is compromised, leading to a degradation in the NMSE performance of all schemes as the number of paths increases. This is a common challenge faced by all compressed sensing-based algorithms. 
Furthermore, as the number of paths increases, the proposed CSEP structure exhibits increased interference, which leads to more significant performance degradation. Nonetheless, ​due to ​its longer utilizable pilot length and the angular domain projection algorithm, the proposed scheme still outperforms other benchmark methods.

To evaluate the computational complexity, Fig.~\ref{CPUTimevsPathNum} presents the CPU running time of different algorithms versus the number of channel propagation paths. In comparison to the SBL method and the 3D-SOMP algorithm, the proposed algorithm for the two pilot structures is computationally more efficient. 
Among them, the higher complexity of the CSEP structure stems from the use of a larger measurement matrix in the compressed sensing framework.
Nevertheless, the CSEP structure outperforms other algorithms in terms of NMSE, BER, and SE. 
This result indicates that it achieves a superior trade-off among complexity, pilot overhead, and estimation accuracy.

\subsection{Performance Comparison with OFDM}

Fig.~\ref{OTFSvsOFDM} shows the performance comparison of multi-user MIMO OTFS systems and multi-user MIMO OFDM systems for different maximum MS velocities, while the other parameters are as the same as Table \ref{simparameters}. The pilot scheme for the OTFS system is the proposed CSEP structure. As shown in Fig.~\ref{OTFSvsOFDM_BER}, OTFS employing the proposed CSEP structure achieves significantly lower BER than OFDM at maximum MS velocities of 100, 200 and 300 km/h. For OFDM, higher velocities exacerbate ICI, consequently degrading the BER performance. For OTFS, it experiences a noticeable performance degradation as the maximum MS speed increases from 100 km/h to 200 km/h. In contrast, a further increase in speed from 200 km/h to 300 km/h leads to only a marginal performance loss. Fig.~\ref{OTFSvsOFDM_SE} shows the SE for OTFS and OFDM. It can be observed that OFDM shows improved SE performance compared to OTFS at low SNR. Once the SNR exceeds 50 dB, the SE performance of OTFS surpasses that of OFDM. Due to severe ICI, the SE of OFDM systems increases with SNR at a slower rate than that of OTFS systems. It can be concluded that OTFS plays a pivotal role and demonstrates superior performance in high-speed mobile communications.

\section{Conclusion}\label{conclusion}
In this paper, we investigated the uplink  channel estimation and pilot design for multi-user MIMO-OTFS systems. 
We proposed a multi-dimensional decomposition-based channel estimation algorithm. The proposed algorithm first estimates the angles via the subspace decomposition-based method. Then,  the received signals are projected onto individual multipath components using acquired angular information, which effectively mitigated the inter-path interference. The fractional delay and Doppler are then estimated via the CS-based off-grid method.
To reduce the pilot overhead in multi-user scenarios, we proposed a novel CSEP pilot structure that leverages the 2D convolution property of the OTFS input-output relationship and the cyclic shift-orthogonality of ZC sequences. 
Compared to the approach of allocating individual pilot regions to different users, the proposed structure significantly reduces the overhead from guard intervals, thereby enhancing the utilization efficiency of the delay-Doppler resource grid. 
Finally, an improved channel estimation algorithm for CSEP pilot structure was proposed.
Simulation results demonstrated that the proposed CSEP scheme and channel estimation algorithm achieve a favorable balance between computational complexity, pilot overhead, channel estimation accuracy, and BER performance.

\begin{figure*}[hb]
	\hrulefill
	\begin{equation}\label{Ytf}
		\begin{aligned}
			Y\left( t,f,n_{\mathrm{BS}} \right) &=\sum_{u=1}^K{}\sum_{n^\prime =0}^{N-1}{\sum_{m^\prime =0}^{M-1}{}}X_{u}^{\mathrm{TF}}\left[ n^{\prime},m^{\prime} \right] \iiint{}\left[ \mathbf{a}\left( \theta _{u,i} \right) \right] _{n_{\mathrm{BS}}}h_u\left( \tau ,\nu ,\theta \right) 
			\\
			&\quad\times \left[ \int{e^{-j2\pi f\left( t\prime -t-T_{\mathrm{CP}} \right)}g_{\mathrm{rx}}^{\ast}\left( t^{\prime}-t \right)}e^{j2\pi \nu \left( t\prime -\tau \right)}e^{j2\pi m\prime \Delta f\left( t\prime -\tau -T_{\mathrm{CP}}-n\prime T_{\mathrm{sym}} \right)}g_{\mathrm{tx}}\left( t^{\prime}-\tau -n^{\prime}T_{\mathrm{sym}} \right) \mathrm{d}t^{\prime} \right] \mathrm{d}\tau \mathrm{d}\nu \mathrm{d}\theta 
			\\
			&=\sum_{u=1}^K{}\sum_{n^\prime =0}^{N-1}{\sum_{m^\prime =0}^{M-1}{}}X_{u}^{\mathrm{TF}}\left[ n^{\prime},m^{\prime} \right] \sum_{i=0}^{P-1}{}\alpha _{u,i}\left[ \mathbf{a}\left( \theta _{u,i} \right) \right] _{n_{\mathrm{BS}}}e^{j2\pi \nu _{u,i}\left( t+T_{\mathrm{CP}}-\tau _{u,i} \right)}e^{j2\pi m\prime \Delta f\left( t-\tau _{u,i}-n^{\prime}T_{\mathrm{sym}} \right)}
			\\
			&\quad\times \int_{t+T_{\mathrm{CP}}}^{t+T_{\mathrm{sym}}}{}e^{-j2\pi \left( f-\nu _{u,i}-m^{\prime}\Delta f \right) \left( t^{\prime}-t-T_{\mathrm{CP}} \right)}g_{\mathrm{rx}}^{\ast}\left( t^{\prime}-t \right) g_{\mathrm{tx}}\left( t^{\prime}-\tau _{u,i}-n^{\prime}T_{\mathrm{sym}} \right) \mathrm{d}t^{\prime}
		\end{aligned}
	\end{equation}
	\hrulefill
		\setcounter{equation}{51}
	\begin{equation}\label{YTFSappendix}
		\begin{aligned}
			Y^{\mathrm{TFS}}\left[ n,m,n_{\mathrm{BS}} \right] &=\sum_{u=1}^K{}\sum_{n^\prime =0}^{N-1}{\sum_{m^\prime =0}^{M-1}{X^{\mathrm{TF}}\left[ n^{\prime},m^{\prime},n_{\mathrm{BS}} \right]}}\sum_{i=0}^{P-1}{}\alpha _{u,i}\left[ \mathbf{a}_{N_{\mathrm{BS}}}\left( \theta _{u,i} \right) \right] _{n_{\mathrm{BS}}}e^{j2\pi \nu _{u,i}\left( nT_{\mathrm{sym}}+T_{\mathrm{CP}}-\tau _{u,i} \right)}
			\\
			&\quad\times e^{j2\pi m^\prime \Delta f\left( nT_{\mathrm{sym}}-\tau _{u,i}-n^{\prime}T_{\mathrm{sym}} \right)}\frac{1}{M}\sum_{p=n\left( M+M_{\mathrm{CP}} \right) +M_{\mathrm{CP}}+l_{u,i}}^{\left( n+1 \right) \left( M+M_{\mathrm{CP}} \right) -1}{}e^{-j2\pi \left( \left( m-m^{\prime} \right) \Delta f-\nu _{u,i} \right) \left( p\frac{T}{M}-n^{\prime}T_{\mathrm{sym}}-T_{\mathrm{CP}} \right)}
			\\
			&=\sum_{u=1}^K{}\sum_{m^\prime =0}^{M-1}{X^{\mathrm{TF}}\left[ n^{\prime},m^{\prime},n_{\mathrm{BS}} \right]}\sum_{i=0}^{P-1}{}\alpha _{u,i}\left[ \mathbf{a}_{N_{\mathrm{BS}}}\left( \theta _{u,i} \right) \right] _{n_{\mathrm{BS}}}e^{j2\pi \nu _{u,i}\left( nT_{\mathrm{sym}}+T_{\mathrm{CP}}-\tau _{u,i} \right)}e^{-j2\pi m^{\prime}\Delta f\tau _{u,i}}
			\\
			&\quad\times A_{g_{\mathrm{rx}},g_{\mathrm{tx}}}^{\prime}\left( -\tau _{u,i},\left( m-m^{\prime} \right) \Delta f-\nu _{u,i} \right).
		\end{aligned}
	\end{equation}
	\hrulefill

	\begin{equation}\label{YDDSappendix}
		\begin{aligned}
			Y^{\mathrm{DDS}}\left[ k,l,n_{\mathrm{BS}} \right] &=\frac{1}{NM}\sum_{u=1}^K{}\sum_{k=-\lceil N/2 \rceil}^{\lceil N/2 \rceil}{\sum_{l^{\prime}=0}^{M-1}{X_{u}^{\mathrm{DD}}[k^{\prime},l^{\prime}]\frac{1}{NM}\sum_{n=0}^{N-1}{\sum_{m=0}^{M-1}{\sum_{m^{\prime}=0}^{M-1}{e^{j2\pi \frac{\left( m-m^{\prime} \right)}{M}M_{\mathrm{CP}}}}}}}\quad}
			\\
			&\quad\times \sum_{i=0}^{P-1}{\alpha _{u,i} \left[ \mathbf{a}_{N_{\mathrm{BS}}}\left( \theta _{u,i} \right) \right] _{n_{\mathrm{BS}}}e^{-j2\pi \left( \nu _{u,i}\left( \tau _{u,i}-nT_{\mathrm{sym}} \right) -m^{\prime}\Delta f\tau _{u,i} \right)}}A_{g_{\mathrm{rx}},g_{\mathrm{tx}}}^{\prime}\left( -\tau _{u,i},\left( m-m^{\prime} \right) \Delta f-\nu _{u,i} \right) 
			\\
			&\quad\times e^{-\jmath 2\pi \left( \frac{m^{\prime}l^{\prime}}{M}-\frac{nk^{\prime}}{N} \right)}e^{-j2\pi \left( \frac{nk}{N}-\frac{ml}{M} \right)}+Z^{\mathrm{DDS}}\left[ k,l,n_{\mathrm{BS}} \right] 
			\\
			&=\sum_{u=1}^K{}\sum_{i=0}^{P-1}{}\sum_{k^{\prime}=-\lceil N/2 \rceil}^{\lceil N/2 \rceil}{\sum_{l^{\prime}=0}^{M-1}{}}\Phi _{u,i}\left( l \right) h_{u,i}\left[ k^{\prime},l^{\prime},n_{\mathrm{BS}} \right] X_{u}^{\mathrm{DD}}\left[ \left[ k-k^{\prime} \right] _N,\left[ l-l^{\prime} \right] _M \right] + Z^{\mathrm{DDS}}\left[ k,l,n_{\mathrm{BS}} \right] .
		\end{aligned}
	\end{equation}
\end{figure*}

\begin{figure*}[t]
	
	\begin{equation}\label{usitj}
		\begin{aligned}
			\mu _{s,i;t,j}&=\left| \left( \mathbf{c}_{t,j}^{\mathrm{peak}} \right) ^{\mathrm{H}}\mathbf{c}_{s,i}^{\mathrm{peak}} \right|
			\\
			&=\left| \sum_l{}\frac{1}{KM_p}e^{j2\pi \frac{\left( k_{s,i}+\kappa _{s,i} \right) \left( M_{\mathrm{CP}}+l-l_{s,i}-\iota _{s,i} \right)}{N\left( M+M_{\mathrm{CP}} \right)}}e^{j\frac{\pi \left( \left[ l-l_{s,i}-l_p \right] _{KM_p} \right)}{KM_p} ^2}e^{-j2\pi \frac{\left( k_{t,j}+\kappa _{t,j} \right) \left( M_{\mathrm{CP}}+l-l_{t,j}-\iota _{t,j} \right)}{N\left( M+M_{\mathrm{CP}} \right)}}e^{-j\frac{\pi\left( \left[ l-l_{t,j}-l_p \right] _{KM_p} \right)}{KM_p} ^2}\right.
			\\
			&\quad\left.\times\sum_k{}\frac{1}{N_p}e^{j\frac{\pi}{N_p}\left( \left[ k-k_{s,i}-k_p \right] _{N_p} \right) ^2}e^{-j\frac{\pi}{N_p}\left( \left[ k-k_{t,j}-k_p \right] _{N_p} \right) ^2} \right|
			\\
			&=\left| \frac{\sin \frac{\pi KM_p\left( k_{s,i}+\kappa _{s,i}-k_{t,j}-\kappa _{t,j} \right)}{N\left( M+M_{\mathrm{CP}} \right)}}{KM_p\sin \pi \left( \frac{\left( k_{s,i}+\kappa _{s,i}-k_{t,j}-\kappa _{t,j} \right)}{N\left( M+M_{\mathrm{CP}} \right)}+\frac{\left( l_{t,j}-l_{s,i} \right) +\left( l_{ps}-l_{pt} \right)}{KM_p} \right)} \right|
		\end{aligned}
	\end{equation}
	\hrulefill
\end{figure*}

\begin{appendices}
	
\section{Proof of Theorem 1}\label{proofoftheorem1}

By substituting (\ref{XTFu}), (\ref{sut}), (\ref{rt}) and (\ref{hDDu}) into (\ref{Atf}), we have $Y\left( t,f,n_{\mathrm{BS}} \right) $ as (\ref{Ytf}), which is at the bottom of the next page. Then, the sampled TFS domain symbols $Y^{\mathrm{TFS}}\left[ n,m,n_{\mathrm{BS}} \right] $ can be obtained by substituting $t=nT_{\mathrm{sym}}, f=m\Delta f$ into (\ref{Ytf}), as shown in (\ref{YTFSappendix}), which is at the bottom of the next page, where the correlation function $A^\prime _{g_{\mathrm{rx}},g_{\mathrm{tx}}}\left( -\tau ,\left( m-m^\prime \right) \Delta f-\nu \right) $ is given by
\setcounter{equation}{50}
\begin{equation}\label{Agtxgrx}
	\begin{aligned}
		&A^\prime _{g_{\mathrm{rx}},g_{\mathrm{tx}}}\left( -\tau ,\left( m-m^\prime \right) \Delta f-\nu \right) 
		\\
		&\quad=\frac{1}{M}\sum_{p=M_{\mathrm{CP}}-M\Delta f\tau}^{M_{\mathrm{CP}}+M-M\Delta f\tau}{e^{-j2\pi \left( \left( m-m^\prime \right) \Delta f-\nu \right) \left( p\frac{T}{M}+\tau \right)}}.
	\end{aligned}
\end{equation}
Next, by substituting (\ref{Agtxgrx}) and (\ref{YTFSappendix}) into (\ref{YDDderivation}), we have the DDS domain symbols $Y^{\mathrm{DDS}}\left[ k,l,n_{\mathrm{BS}} \right] $ as shown in (\ref{YDDSappendix}), which is at the bottom of the next page, which completes the proof.

\section{Proof of Theorem 2}\label{proofoftheorem2}
According to (\ref{c_si}) and (\ref{c_tj}), the coherence $\mu _{s,i;t,j}$ is given in (\ref{usitj}), which is at the top of the page after next. 
Based on the properties of the function $\frac{\sin \pi x}{\sin \frac{\pi}{X}x}$, when $\left( k_{s,i}+\kappa _{s,i}-k_{t,j}-\kappa _{t,j} \right) $ and $\left( l_{t,j}-l_{s,i} \right) +\left( l_{ps}-l_{pt} \right) $ have the same sign, $\frac{\left( k_{s,i}+\kappa _{s,i}-k_{t,j}-\kappa _{t,j} \right)}{N\left( M+M_{\mathrm{CP}} \right)}+\frac{\left( l_{t,j}-l_{s,i} \right) +\left( l_{ps}-l_{pt} \right)}{KM_p}$ is sufficiently far from zero, and the resulting coherence is small. 
In contrast, the coherence is high when $\left( k_{s,i}+\kappa _{s,i}-k_{t,j}-\kappa _{t,j} \right) $ and $\left( l_{t,j}-l_{s,i} \right) +\left( l_{ps}-l_{pt} \right) $ have the opposite sign as $\frac{\left( k_{s,i}+\kappa _{s,i}-k_{t,j}-\kappa _{t,j} \right)}{N\left( M+M_{\mathrm{CP}} \right)}+\frac{\left( l_{t,j}-l_{s,i} \right) +\left( l_{ps}-l_{pt} \right)}{KM_p}$ is close to zero. 

Next, we discuss the conditions under which $\mu _{s,i;t,j}$ achieves its maximum value.
First, the term $\frac{\left( k_{s,i}+\kappa _{s,i}-k_{t,j}-\kappa _{t,j} \right)}{N\left( M+M_{\mathrm{CP}} \right)}$ lies within the range $\left[ -\frac{N_g}{N\left( M+M_{\mathrm{CP}} \right)},\frac{N_g}{N\left( M+M_{\mathrm{CP}} \right)} \right] $. Since $N_g<N$ and $KM_p<M$, we have 
$$
-\frac{1}{M_p}<\frac{\left( k_{s,i}+\kappa _{s,i}-k_{t,j}-\kappa _{t,j} \right)}{N\left( M+M_{\mathrm{CP}} \right)}<\frac{1}{M_p}.
$$
Second, the sign of the term $\frac{\left( l_{t,j}-l_{s,i} \right) +\left( l_{ps}-l_{pt} \right)}{KM_p}$ is always the same as the sign of $\left( l_{ps}-l_{pt} \right) $, since $\left( l_{t,j}-l_{s,i} \right) \in \left[ -M_g,M_g \right] $ and $M_g<M_p$, and $\left( l_{ps}-l_{pt} \right) =\left( s-t \right) M_p$. When $\left( l_{t,j}-l_{s,i} \right) =-M_g$ and $s-t=1$, we have 
$$
\frac{\left( l_{t,j}-l_{s,i} \right) +\left( l_{ps}-l_{pt} \right)}{KM_p}=\frac{M_p-M_g}{KM_p},
$$
which is closest to 0, resulting in maximum coherence $\mu _{s,i;t,j}$. It can be observed that the coherence decreases as $M_p$ increases. Considering the minimal value of $M_p=M_g+1$, the term $\frac{\left( l_{t,j}-l_{s,i} \right) +\left( l_{ps}-l_{pt} \right)}{KM_p}$ reaches its minimum value of $1/KM_p$.

Based on the above discussion, coherence $\mu _{s,i;t,j}$ is maximized when $\left( k_{s,i}+\kappa _{s,i}-k_{t,j}-\kappa _{t,j} \right) =N_g$ and $\left( l_{t,j}-l_{s,i} \right) +\left( l_{ps}-l_{pt} \right) =1$, and its maximum value is given by
\setcounter{equation}{54}
\begin{equation}\label{maxMusitj}
	\mu _{s,i;t,j}^{\max}=\left| \frac{\sin \frac{\pi KM_pN_g}{N\left( M+M_{\mathrm{CP}} \right)}}{KM_p\sin \pi \left( \frac{N_g}{N\left( M+M_{\mathrm{CP}} \right)}-\frac{1}{KM_p} \right)} \right|.
\end{equation}
To investigate the relation ship between the size of $KM_p$ and the maximum coherence $\mu _{s,i;t,j}^{\max}$, we set $\frac{\left( k_{s,i}+\kappa _{s,i}-k_{t,j}-\kappa _{t,j} \right)}{N\left( M+M_{\mathrm{CP}} \right)}=\frac{a}{KM_p}$, $0<a<1$. Then, (\ref{maxMusitj}) can be rewritten as 
\begin{equation}
	\begin{aligned}
\mu _{s,i;t,j}^{\max}&=\left| \frac{\sin \pi a}{KM_p\sin  \frac{\pi(a-1)}{KM_p}} \right|
\\
&\overset{a}{\approx}\left| \frac{ a}{ {a-1}} \right|,
	\end{aligned}
\end{equation}
where the approximation $a$ is due to the fact that $\sin x\approx x$ when $x$ is close to 0. Assume that $\mu _{s,i;t,j}$ does not exceed a certain threshold $\epsilon \ll 1$, we have $a\le \frac{\epsilon}{1+\epsilon}$. Under this condition, $KM_p$ satisfies
\begin{equation}
	KM_p<\frac{\epsilon}{1+\epsilon}\frac{N\left( M+M_{\mathrm{CP}} \right)}{N_g}\approx \frac{\epsilon N\left( M+M_{\mathrm{CP}} \right)}{N_g},
\end{equation}
which completes the proof.

\end{appendices}

\bibliographystyle{IEEEtran}
\bibliography{myre}

\begin{thebibliography}{10}
\providecommand{\url}[1]{#1}
\csname url@samestyle\endcsname
\providecommand{\newblock}{\relax}
\providecommand{\bibinfo}[2]{#2}
\providecommand{\BIBentrySTDinterwordspacing}{\spaceskip=0pt\relax}
\providecommand{\BIBentryALTinterwordstretchfactor}{4}
\providecommand{\BIBentryALTinterwordspacing}{\spaceskip=\fontdimen2\font plus
\BIBentryALTinterwordstretchfactor\fontdimen3\font minus \fontdimen4\font\relax}
\providecommand{\BIBforeignlanguage}[2]{{%
\expandafter\ifx\csname l@#1\endcsname\relax
\typeout{** WARNING: IEEEtran.bst: No hyphenation pattern has been}%
\typeout{** loaded for the language `#1'. Using the pattern for}%
\typeout{** the default language instead.}%
\else
\language=\csname l@#1\endcsname
\fi
#2}}
\providecommand{\BIBdecl}{\relax}
\BIBdecl

\bibitem{6g}
W.~Jiang, B.~Han, M.~A. Habibi, and H.~D. Schotten, ``The road towards 6g: A comprehensive survey,'' \emph{IEEE Open Journal of the Communications Society}, vol.~2, pp. 334--366, 2021.

\bibitem{otfstutorial}
Z.~Wei, W.~Yuan, S.~Li, J.~Yuan, G.~Bharatula, R.~Hadani, and L.~Hanzo, ``Orthogonal time-frequency space modulation: A promising next-generation waveform,'' \emph{IEEE Wireless Commun.}, vol.~28, no.~4, pp. 136--144, Aug. 2021.

\bibitem{otfs}
R.~Hadani, S.~Rakib, M.~Tsatsanis, A.~Monk, A.~J. Goldsmith, A.~F. Molisch, and R.~Calderbank, ``Orthogonal time frequency space modulation,'' in \emph{2017 IEEE Wireless Communications and Networking Conference (WCNC)}, 2017, pp. 1--6.

\bibitem{otfs2}
\BIBentryALTinterwordspacing
R.~Hadani and A.~Monk, ``Otfs: A new generation of modulation addressing the challenges of 5g,'' 2018. [Online]. Available: \url{https://arxiv.org/abs/1802.02623}
\BIBentrySTDinterwordspacing

\bibitem{ofdmotfs}
A.~Farhang, A.~RezazadehReyhani, L.~E. Doyle, and B.~Farhang-Boroujeny, ``Low complexity modem structure for {OFDM}-based orthogonal time frequency space modulation,'' \emph{IEEE Wireless Communications Letters}, vol.~7, no.~3, pp. 344--347, 2018.

\bibitem{embeddedpilot}
P.~Raviteja, K.~T. Phan, and Y.~Hong, ``Embedded pilot-aided channel estimation for {OTFS} in delay–doppler channels,'' \emph{IEEE Trans. Veh. Tech.}, vol.~68, no.~5, pp. 4906--4917, May 2019.

\bibitem{sblotfs}
Z.~Wei, W.~Yuan, S.~Li, J.~Yuan, and D.~W.~K. Ng, ``Off-grid channel estimation with sparse bayesian learning for {OTFS} systems,'' \emph{IEEE Trans. Wireless Commun.}, vol.~21, no.~9, pp. 7407--7426, Mar. 2022.

\bibitem{sblotfs3}
M.~Tang, H.~Wang, Z.~Yuan, and J.~Yuan, ``A novel off-grid channel estimation with fast bcs using lsm prior for otfs modulation,'' \emph{IEEE Trans. Wireless Commun.}, vol.~23, no.~9, pp. 12\,157--12\,171, Sep 2024.

\bibitem{shan}
Y.~Shan, F.~Wang, Y.~Hao, J.~Yuan, J.~Hua, and Y.~Xin, ``Off-grid channel estimation using grid evolution for {OTFS} systems,'' \emph{IEEE Trans. Wireless Commun.}, vol.~23, no.~8, pp. 9549--9565, Aug. 2024.

\bibitem{shen}
W.~Shen, L.~Dai, J.~An, P.~Fan, and R.~W. Heath, ``Channel estimation for orthogonal time frequency space ({OTFS}) massive {MIMO},'' \emph{IEEE Trans. Signal Process.}, vol.~67, no.~16, pp. 4204--4217, Aug. 2019.

\bibitem{dingshi}
D.~Shi, W.~Wang, L.~You, X.~Song, Y.~Hong, X.~Gao, and G.~Fettweis, ``Deterministic pilot design and channel estimation for downlink massive {MIMO-OTFS} systems in presence of the fractional doppler,'' \emph{IEEE Trans. Wireless Commun.}, vol.~20, no.~11, pp. 7151--7165, Nov. 2021.

\bibitem{multidim}
A.~Mehrotra, J.~Singh, S.~Srivastava, R.~Kumar~Singh, A.~K. Jagannatham, and L.~Hanzo, ``Multi-dimensional sparse csi acquisition for hybrid mmwave {MIMO OTFS} systems,'' \emph{IEEE Trans. Commun.}, vol.~73, no.~9, pp. 8330--8344, Mar. 2025.

\bibitem{tfmultiaccess}
R.~M. Augustine and A.~Chockalingam, ``Interleaved time-frequency multiple access using otfs modulation,'' in \emph{2019 IEEE 90th Vehicular Technology Conference (VTC2019-Fall)}, 2019, pp. 1--5.

\bibitem{multiaccess}
\BIBentryALTinterwordspacing
G.~D. Surabhi, R.~M. Augustine, and A.~Chockalingam, ``Multiple access in the delay-doppler domain using {OTFS} modulation,'' 2019. [Online]. Available: \url{https://arxiv.org/abs/1902.03415}
\BIBentrySTDinterwordspacing

\bibitem{multiaccess5}
M.~Nie, S.~Li, D.~Mishra, J.~Yuan, and D.~Wing Kwan~Ng, ``Uplink multi-user otfs: Transmitter design based on statistical channel information,'' \emph{IEEE Trans. Commun.}, vol.~73, no.~7, pp. 4678--4696, Jul. 2025.

\bibitem{multiaccess7}
V.~Khammammetti and S.~K. Mohammed, ``Spectral efficiency of otfs based orthogonal multiple access with rectangular pulses,'' \emph{IEEE Trans. Vehi. Technol.}, vol.~71, no.~12, pp. 12\,989--13\,006, Aug. 2022.

\bibitem{multiaccess2}
M.~Bayat, S.~P.S., and A.~Farhang, ``Time and frequency synchronization for multiuser otfs in uplink,'' \emph{IEEE Trans. Vehi. Technol.}, pp. 1--17, Oct. 2025.

\bibitem{multiaccess6}
V.~Khammammetti and S.~K. Mohammed, ``Otfs-based multiple-access in high doppler and delay spread wireless channels,'' \emph{IEEE Wireless Commun. Lett.}, vol.~8, no.~2, pp. 528--531, Apr. 2019.

\bibitem{multiaccess3}
M.~Li, S.~Zhang, F.~Gao, P.~Fan, and O.~A. Dobre, ``A new path division multiple access for the massive {MIMO-OTFS} networks,'' \emph{IEEE J. Sel. Areas Commun.}, vol.~39, no.~4, pp. 903--918, Apr. 2021.

\bibitem{multiaccess4}
M.~Zhou, F.~Chen, M.~Xia, X.~Zhang, and H.~Yu, ``Iterative channel estimation for multi-user otfs uplink systems with superimposed full pilots,'' \emph{IEEE Trans. Vehi. Technol.}, vol.~74, no.~3, pp. 4485--4497, Mar. 2025.

\bibitem{jakes1994microwave}
W.~C. Jakes and D.~C. Cox, \emph{Microwave mobile communications}.\hskip 1em plus 0.5em minus 0.4em\relax New York, NY, USA: Wiley, 1994.

\bibitem{raviteja}
P.~Raviteja, K.~T. Phan, Y.~Hong, and E.~Viterbo, ``Interference cancellation and iterative detection for orthogonal time frequency space modulation,'' \emph{IEEE Trans. Wireless Commun.}, vol.~17, no.~10, pp. 6501--6515, Oct. 2018.

\bibitem{VBI}
Q.~Wang, M.~Lei, M.-M. Zhao, and M.~Zhao, ``Variational bayesian inference based channel estimation for otfs system with lsm prior,'' in \emph{2022 International Symposium on Wireless Communication Systems (ISWCS)}, 2022, pp. 1--5.

\bibitem{2dsbl}
Q.~Wang, Y.~Liang, Z.~Zhang, and P.~Fan, ``2d off-grid decomposition and sbl combination for otfs channel estimation,'' \emph{IEEE Trans. Wireless Commun.}, vol.~22, no.~5, pp. 3084--3098, May 2023.

\bibitem{3dsomp}
A.~Mehrotra, J.~Singh, S.~Srivastava, R.~Kumar~Singh, A.~K. Jagannatham, and L.~Hanzo, ``Multi-dimensional sparse csi acquisition for hybrid mmwave mimo otfs systems,'' \emph{IEEE Trans. Commun.}, vol.~73, no.~9, pp. 8330--8344, Sep. 2025.

\bibitem{8686339}
G.~D. Surabhi, R.~M. Augustine, and A.~Chockalingam, ``On the diversity of uncoded otfs modulation in doubly-dispersive channels,'' \emph{IEEE Trans. Wireless Commun.}, vol.~18, no.~6, pp. 3049--3063, Jun. 2019.

\end{thebibliography}


\end{document}